\documentclass[twocolumn]{aastex63}

\received{September 16, 2020}
\revised{January 20, 2021}
\accepted{January 20, 2021}

\submitjournal{ApJ}

\shorttitle{Star Formation in M\,36}
\shortauthors{Panja et al. }

\begin{document}


\title{Sustaining Star Formation in the Galactic Star Cluster M\,36?}

\correspondingauthor{Alik Panja}
\email{alik.panja@gmail.com}

\author[0000-0002-4719-3706]{Alik Panja}
\affiliation{S. N. Bose National Centre for Basic Sciences, Kolkata 700106, India}

\author[0000-0003-0262-272X]{Wen Ping Chen}
\affiliation{Graduate Institute of Astronomy, National Central University, 300 Zhongda Road, Zhongli, Taoyuan 32001, Taiwan}

\author[0000-0002-2338-4583]{Somnath Dutta}
\affiliation{Institute of Astronomy and Astrophysics, Academia Sinica, Taipei 10617, Taiwan}

\author[0000-0002-3904-1622]{Yan Sun}
\affiliation{Purple Mountain Observatory, Chinese Academy of Sciences, 10 Yuanhua Road, Nanjing 210033, China}

\author[0000-0003-0007-2197]{Yu Gao} 
\affiliation{Purple Mountain Observatory, Chinese Academy of Sciences, 10 Yuanhua Road, Nanjing 210033, China}
\affiliation{Department of Astronomy, Xiamen University, Xiamen, Fujian 361005, China}

\author[0000-0003-1457-0541]{Soumen Mondal}
\affiliation{S. N. Bose National Centre for Basic Sciences, Kolkata 700106, India}

\begin{abstract}

We present comprehensive characterization of the Galactic open cluster M\,36.  
Some two hundred member candidates, with an estimated contamination rate of $\sim8$\%, have been identified on the basis of proper motion and parallax measured by the {\it Gaia}~DR2.  The cluster has a proper motion grouping around ($\mu_{\alpha} \cos\delta = -0.15 \pm 0.01$~mas~yr$^{-1}$, and $\mu_{\delta} = -3.35 \pm 0.02$~mas~yr$^{-1}$), distinctly separated from the field population.  Most member candidates have parallax values 0.7--0.9~mas, with a median value of 0.82 $\pm$ 0.07~mas (distance $\sim1.20 \pm 0.13$~kpc).  The angular diameter of $27\arcmin \pm 0\farcm4$ determined from the radial density profile then corresponds to a linear extent of $9.42 \pm0.14$~pc.  
With an estimated age of $\sim15$~Myr, M\,36 is free of nebulosity.  To the south-west of the cluster, we discover a highly obscured ($A_{V}$ up to $\sim23$~mag), compact ($\sim 1\farcm9 \times 1\farcm2$) dense cloud, within which three young stellar objects in their infancy (ages $\la 0.2$~Myr) are identified.  The molecular gas, 3.6~pc in extent, contains a total mass of (2--3)$\times10^{2}~M_{\sun}$, and has a uniform velocity continuity across the cloud, with a velocity range of $-20$ to $-22$~km~s$^{-1}$, consistent with the radial velocities of known star members.  
In addition, the cloud has a derived kinematic distance marginally in agreement with that of the star cluster.  If physical association between M\,36 and the young stellar population can be unambiguously established, this manifests a convincing example of prolonged star formation activity spanning up to tens of Myrs in molecular clouds.

\end{abstract}

\section{Introduction}
 \label{sec:intro}

A giant molecular cloud fragments to form aggregates of stellar associations or clusters.  As the giant molecular cloud itself disperses, individual aggregates may or may not remain gravitationally bound \citep{lad03,kru19}.  ``Timescale'' is key to the hierarchical structuring.  
Giant molecular clouds have a life expectancy of $\sim10^{8}$~years in orbiting the Galactic plane through spiral arms.  More massive ones ($\gtrsim10^{5}$~M$_\sun$), capable of producing thousands of stars, thereby likely hosting luminous OB stars, are more vulnerable to photoevaporation, and would be destructed on timescales $\lesssim 30$--40~Myr \citep{wil97}.  
Within a cloud that ends up producing a cluster, starbirth may not proceed simultaneously in the dense cores hosting individual protostars.  With a coeval onset of formation, massive stars will be gone by a couple of million years, whereas low-mass members in the same star cluster would take tens of million years to evolve to the main sequence stage.    
  
Stellar formation and evolution is hardly in isolation, however.  The spatial structure of an open cluster starts out as the relic of the conditions in the parental cloud; that is, massive stars may preferentially be formed in the denser, central parts of a cloud, and via their strong radiation and stellar winds, disrupt the surrounding gas and dust, hindering any subsequent star formation activity.  Alternatively, given proper conditions, the ionizing or explosive shocks may compress neighboring clouds, hence triggering the next epoch of star formation.  In this scenario, an evolved stellar population and an episodically younger group of stars would coexist in the same region.

M\,36 (NGC\,1960, R.A.~$= 05^{\rm h}36^{\rm m}18^{\rm s}$, Decl.~$= +34\degr08\arcmin24\arcsec$, J2000) is a rich star cluster near the center of the Aur~OB1 association.  Dominated by about 15 bright ($V\lesssim 11$~mag) stars, the cluster spans an angular extent of $\sim10\arcmin$ \citep{lyn87}.   
\citet{bar85} estimated a reddening $E(B-V)=0.24$~mag, a distance of 1.20~kpc to the cluster, and an age of 30~Myr from the main-sequence turn-off. 
With proper motion data (to a limit of $V \sim 14$~mag), $BV$ photometry ($V \lesssim 19$~mag), and using isochrone fitting, \citet{san00} derived $E(B-V)=0.25 \pm 0.02$~mag, a distance of $1.32\pm 0.12$~kpc, an age of $16_{-5}^{+10}$~Myr, and metallicity of $Z = 0.02$.  
\citet{hai01} reported a circumstellar disk fraction of $3\% \pm 3\%$ on the basis of $JHKL$ photometry, suggestive a cluster age of 30~Myr. \citet{sha06} estimated a core radius to be $3\farcm2$ (1.2~pc), extending up to $14 \arcmin$ (5.4~pc) using the projected radial density profile of the main-sequence stars down to $V \lesssim 18$~mag, and concluded that the core of this relatively young cluster is spherical, whereas the outer region is experiencing external disturbances.  
Furthermore, from the optical and infrared color-color and color-magnitude diagram analysis, they calculated $\log ({\rm age}) = 7.4$ ($25$~Myr) and a distance of $\sim 1.33$~kpc, with a reddening $E(B-V) = 0.22$~mag.  
Using \citet{joh53} photometry, \citet{may08} reported $E(B-V)=0.20 \pm 0.02$~mag, a distance of $1.17_{-0.04}^{+0.06}$~kpc, and an age of $\sim20$~Myr. \citet{bel13}, on the other hand, derived an age of 20~Myr, a distance modulus of 10.33$_{-0.05}^{+0.02}$~mag (hence a distance of 1.16$_{-0.03}^{+0.01}$~kpc), and $E(B-V)=0.20$~mag. 
Applying the lithium depletion boundary technique on a sample of very low-mass cluster members, \citet{jef13} determined an age of $22\pm4$~Myr.  

So far, literature results have indicated consistently an age of 20--30~Myr for M\,36. 
In this paper, we present membership identification using the {\it Gaia}~DR2 data to re-affirm the age and other parameters of the cluster.  Moreover, we recognize a nearby ($\sim 2\farcm8$ to the south-west) stellar population in association with molecular clouds that harbor protostellar activity, i.e., with an age less than a couple of Myr, and is physically related to M\,36, thereby providing a compelling case of prolonged star formation spanning up to 30~Myr in the cloud complex.  
We describe in Section~\ref{sec:data_sets} the data sets used in the study, their quality and limitations, and analysis processes.  The results are presented in Section~\ref{sec:res} on how cluster members are selected, from which the cluster parameters are derived.  
Despite its Messier entry, somewhat surprisingly, the cloud contents in M\,36 have not been well characterized before.  Section~\ref{sec:ext_yso} presents the distribution of extinction, in relevance to the young stellar population in the region.  This leads to the discussion in Section~\ref{sec:dis}, presents molecular cloud structures using CO line emissions and implication of sustaining starbirth processes.  We then summarize the main results of our study in Section~\ref{sec:con}.

\section{Data Sources}
 \label{sec:data_sets}

This study has made use of a variety of data.  Stellar membership is diagnosed on the basis of space position, proper motion, and parallax. A combination of near- to far-infrared photometric data allows us to estimate the extinction distribution and identification of young stellar objects (YSOs).  Molecular line emissions have been measured for CO isotopes to trace the spatial extents and kinematics of dense gas, as well as the possible star formation link of YSOs and molecular gas associated with the cluster, with which some of the member stars with radial velocity measurements are compared.  A brief description of each of these data sets is given as follows.

\subsection{Astrometry and Photometry from the {\it Gaia}~DR2}
 \label{ssec:data_gaia}

The {\it Gaia} Data Release~2 \citep[{\it Gaia}~DR2]{gai18} furnishes precise astrometry on five parameters, i.e., celestial coordinates, trigonometric parallaxes, and proper motions for more than 1.3 billion objects from the observations and analysis of the European Space Agency {\it Gaia} satellite \citep{gai16}. In addition to the homogeneous astrometry over the whole sky, {\it Gaia}~DR2 is further enriched with photometry of three broad-band magnitudes in $G$ (330--1050~nm), $G_{BP}$ (330--680~nm), and $G_{RP}$ (630--1050~nm) with unprecedented accuracy. 
The $G$ band photometry ranges between 3~mag and 21~mag, although stars with $G \lesssim6$~mag generally have inferior astrometry due to calibration issues. The median uncertainty in parallax is about 0.04~mas for bright ($G \la14$~mag) sources, 0.1~mas at $G=17$~mag, and 0.7~mas at $G=20$~mag \citep{lur18}. The corresponding uncertainties in proper motion components are 0.05, 0.2, and 1.2~mas~yr$^{-1}$, respectively \citep{lin18}. The {\it Gaia}~DR2 also provides radial velocity measurements for sources with $G \lesssim13$~mag, and a photometric catalogue of about 0.5 million variable stars \citep{gai18}.

\subsection{Infrared Photometry from the UKIDSS, 2MASS, and {\it Spitzer}}
 \label{ssec:data_infr}

The near-infrared $J$-, $H$-, and $K$-band photometric data were obtained from the UKIDSS DR10PLUS Galactic Plane Survey (GPS; \citealt{law07}) and the 2MASS Point Source Catalog (PSC; \citealt{skr06}). UKIDSS has a finer angular resolution ($0\farcs8$ FWHM, $0\farcs2$ pixels at $K$) compared to 2MASS (1$\arcsec$ FWHM, 2$\arcsec$ pixels), as well as fainter limits. The UKIDSS GPS has median 5$\sigma$ detection limits at $J = 19.77$~mag, $H = 19.00$~mag, and $K = 18.05$~mag \citep{luc08}, whereas 2MASS is limited to $J = 15.8$, $H = 15.1$, and $K_S = 14.3$~mag (S/N = 10; \citealt{skr06}). Selection of reliable sources from the UKIDSS catalog was carried out by utilizing the Structured Query Language (SQL\footnote{\url{http://wsa.roe.ac.uk/sqlcookbook.html}}) query and adopted the criteria published by \citet{luc08}, which eliminates saturated, non-stellar, and unreliable sources near the sensitivity limits. The UKIDSS GPS shows saturation limits at $J = 13.25$~mag, $H = 12.75$~mag, and $K=12.00$~mag \citep{luc08}.  To supplement the UKIDSS saturated sources with 2MASS photometry, we set 0.5~mag fainter than quoted values \citep{ale13, dut15}. For the near-infrared bands, photometric uncertainty $\la0.1$~mag was considered as quality criteria, which give  S/N $\ga 10$.

The mid-infrared photometric data for point sources toward the M\,36 cluster was obtained from the {\it Spitzer} Warm Mission \citep{hor12} survey. Magnitudes for Infrared Array Camera (IRAC; \citealt{faz04}) 
[3.6] \micron\ and [4.5] \micron\ bands were downloaded from the  Glimpse360\footnote{\url{http://www.astro.wisc.edu/sirtf/glimpse360/}} catalog (Program Name/Id: {\small WHITNEY\_GLIMPSE360\_2/61070}) with a pixel scale of 1$\farcs$2~pixel$^{-1}$. We restricted the sources with uncertainty $<$ 0.2~mag for all the IRAC bands to achieve good quality photometric catalog.

\subsection{Stellar Parameters from the LAMOST DR5}
 \label{ssec:data_lamo}

The Large Sky Area Multi-Object Fiber Spectroscopic Telescope (LAMOST, also called the Guoshoujing Telescope) is a reflecting Schmidt telescope \citep{wan96} with an effective aperture of 3.6--4.9~m, a focal length of 20~m, and a field of view of $5\degr$ \citep{cui12, zha12}.
After a five-year regular survey\footnote{\url{http://www.lamost.org/public/node/311?locale=en}} conducted between 2012 to 2017, more than eight million spectra were obtained, with a spectral resolution of R $\sim 1800$ over the wavelength range of 370--900~nm.
The raw 2D spectra are processed uniformly with the LAMOST 2D reduction pipeline \citep{luo15} to generate the 1D spectra, from which stellar parameters, including radial velocity (RV), effective temperature (T$_{eff}$), surface gravity ($\log g$), and metallicity ([Fe/H]), are then derived. Both the 1D spectra and the stellar parameters are publicly available via the LAMOST data releases\footnote{\url{http://www.lamost.org/}} \citep{luo12, luo15}.
The stellar parameters are collected from LAMOST DR5\footnote{\url{http://dr5.lamost.org/}} for this work.

\subsection{CO Data and Reduction}
 \label{ssec:data_pmo}

The CO observations toward M\,36 were conducted on 2 March 2014 using the 13.7~m millimetre-wavelength telescope of the Purple Mountain Observatory~(PMO) in Delingha, China, as a part of the Milky Way Imaging Scroll Painting (MWISP) project dedicated to map the molecular gas along the northern Galactic plane \citep{su19}. The nine-beam Superconducting Spectroscopic Array Receiver system was used as the front end, and each Fast Fourier transform spectrometer with a bandwidth of 1~GHz provided 16,384 channels and a spectral resolution of 61~kHz \citep[see the details in][]{sha12}. The molecular lines of $^{12}$CO ($J$=1--0) in the upper sideband, and $^{13}$CO ($J$=1--0) together with C$^{18}$O ($J$=1--0) in the lower sideband were observed simultaneously. Typical system temperature is 150--200~K in the lower sideband and 250--300~K in the upper sideband. The on-the-fly mode was applied with typical sample steps $10\arcsec$--$15\arcsec$ and scanned along both the Galactic longitude and latitude, in order to reduce the fluctuation of noise. The on-the-fly raw data were then resampled into $30\arcsec \times30\arcsec$ grids and mosaicked to a FITS cube using the GILDAS software package\footnote{\url{http://www.iram.fr/IRAMFR/GILDAS}}.

For the data reported here, the typical sensitivity was about 0.3~K~(T$_{\rm mb}$) in $^{13}$CO and C$^{18}$O at the resolution of 0.17~km~s$^{-1}$, and 0.5~K~(T$_{\rm mb}$) in $^{12}$CO at the resolution of 0.16~km~s$^{-1}$. The beam widths were about $55\arcsec$ and $52\arcsec$ at 110~GHz and 115~GHz, respectively. The pointing of the telescope has an rms accuracy of about $5\arcsec$. It should be noted that any results presented in the figures and tables are on the brightness temperature scale (T$_{\rm R}^{*}$), corrected with the beam efficiencies (from the status report of the telescope\footnote{\url{http://www.radioast.nsdc.cn/mwisp.php}}) using $T_{\rm mb}$ = $T_{\rm A}^{\ast} / \eta_{\rm mb}$, with a calibration accuracy within 10\%.

\section{Results}
 \label{sec:res}

The proper motion, distance, and radial velocity are key parameters to study the fundamental properties of open clusters, as well as the Galactic dynamics \citep{dia05, wu09}. A consistent membership probability can also be estimated from the photometric study of the sources \citep{wu07}. Open star clusters are embedded in the Galactic disk population and identification of membership is generally affected by the field star contamination. The primary source of field stars are foreground or background stars in the direction of the cluster, that have different origins and evolutionary phases. Hence, the membership analysis of stars in a cluster region has become an intense subject of interest to understand the cluster properties and its evolution.

Identification of members within the cluster region relies on grouping of stars in spatial distribution, proper motion distribution, and measurements of parallax. We used a combination of UKIDSS, 2MASS, and {\it Gaia}~DR2 catalogs to estimate those parameters.

\subsection{Radial Density Profile}
 \label{ssec:rad_pro}

M\,36 has been known to have an elongated shape, with the aspect ratio of 0.2--0.3, tilted some $20\degr$ away from the plane \citep{che04, kha09}. Using King profiles, \citet{pis07} estimated an empirical angular radius $16\farcm2$, and a total mass $\sim 200$~M$_\sun$ within the tidal radius of 9~pc.

\begin{figure}
\centering
  \includegraphics[width=1.0\columnwidth,angle=0]{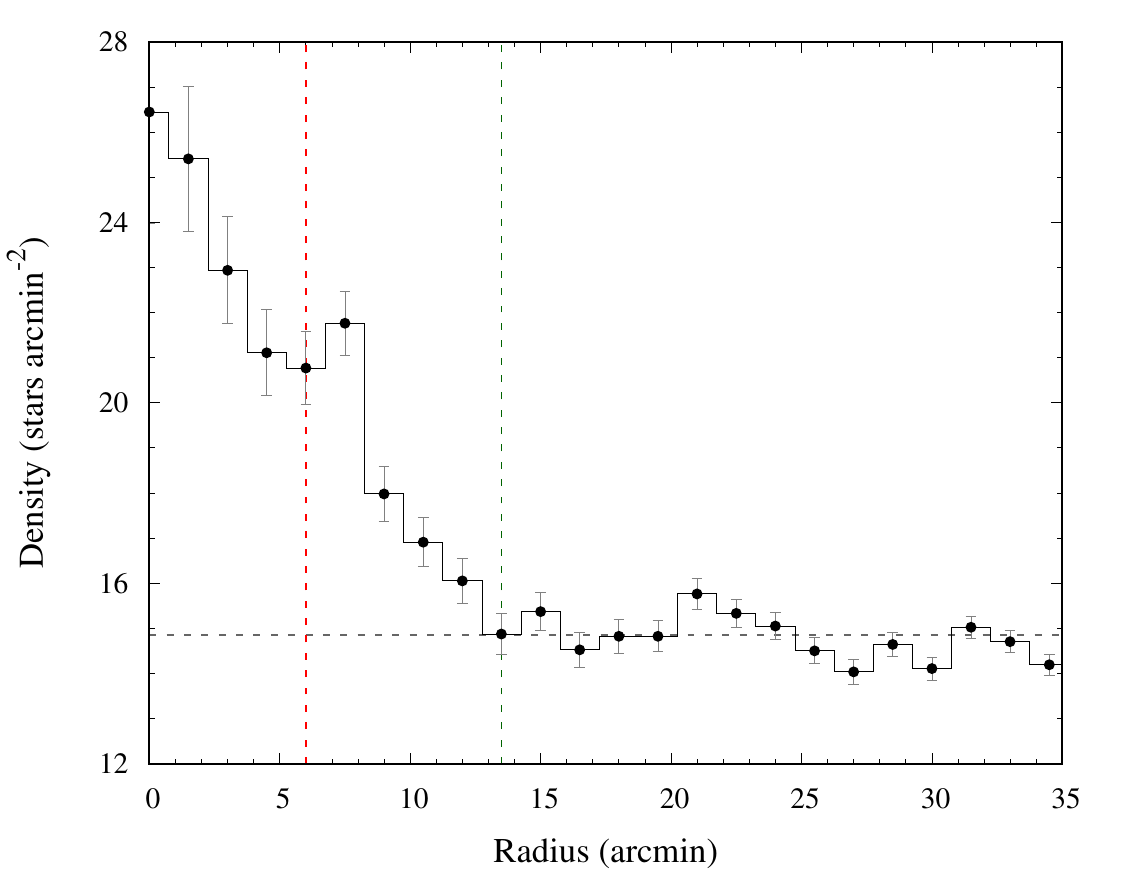}
  \caption{Radial density profile of the stars toward M\,36, taken from the UKIDSS $K$-band data.  The cluster radius is taken as $13\farcm5$, depicted by the green vertical dashed line, where the cluster density merges with that of the background field, marked by the horizontal line. The red vertical line near $6\arcmin$ signifies the fluctuation due to the boundary of the molecular cloud structure (Section~\ref{ssec:mor_par} and \ref{ssec:mol_yso}).}
  \label{fig:rad_den}
\end{figure}

In order to determine the cluster extension, we utilized the UKIDSS $K$-band data. The central coordinates ($\alpha_{2000}$ = $05^{\rm{h}}36^{\rm{m}}18^{\rm{s}}$, $\delta_{2000}$ = $+ 34^{\rm{d}}08^{\rm{m}}24^{\rm{s}}$) as informed in the SIMBAD database\footnote{\url{http://simbad.u-strasbg.fr/simbad/sim-fid}} are adopted to construct the radial density profile with respect to the background. The radial profile is generated by counting the number of stars inside, and divided by the area of, each concentric annulus of 1.5 arcmin width, up to a radius of 35~arcmin. The projected radial stellar density distribution toward M\,36 is shown in Figure~\ref{fig:rad_den}. The peak radial density is estimated to be $\sim 26.4 \pm 2.5$ stars~arcmin$^{-2}$, in contrast to the mean background stellar density of $\sim 14.8 \pm 0.3$ stars~arcmin$^{-2}$.  The density profile starts to blend with that of the field population at radius $\sim 13\farcm5 \pm 0\farcm4$, which we adopt as the cluster boundary.  Our estimation of the cluster radius is consistent with the earlier work of \citet{sha06}, determined as $14\arcmin$ using optical photometry. The fluctuation around radius $6\arcmin$ arises from the possible boundary of the associated molecular cloud (Section~\ref{ssec:mor_par} and \ref{ssec:mol_yso}).

\subsection{Astrometric Membership Criteria}
 \label{ssec:ast_mem}

\begin{figure*}
\centering
  \includegraphics[height=0.9\textheight]{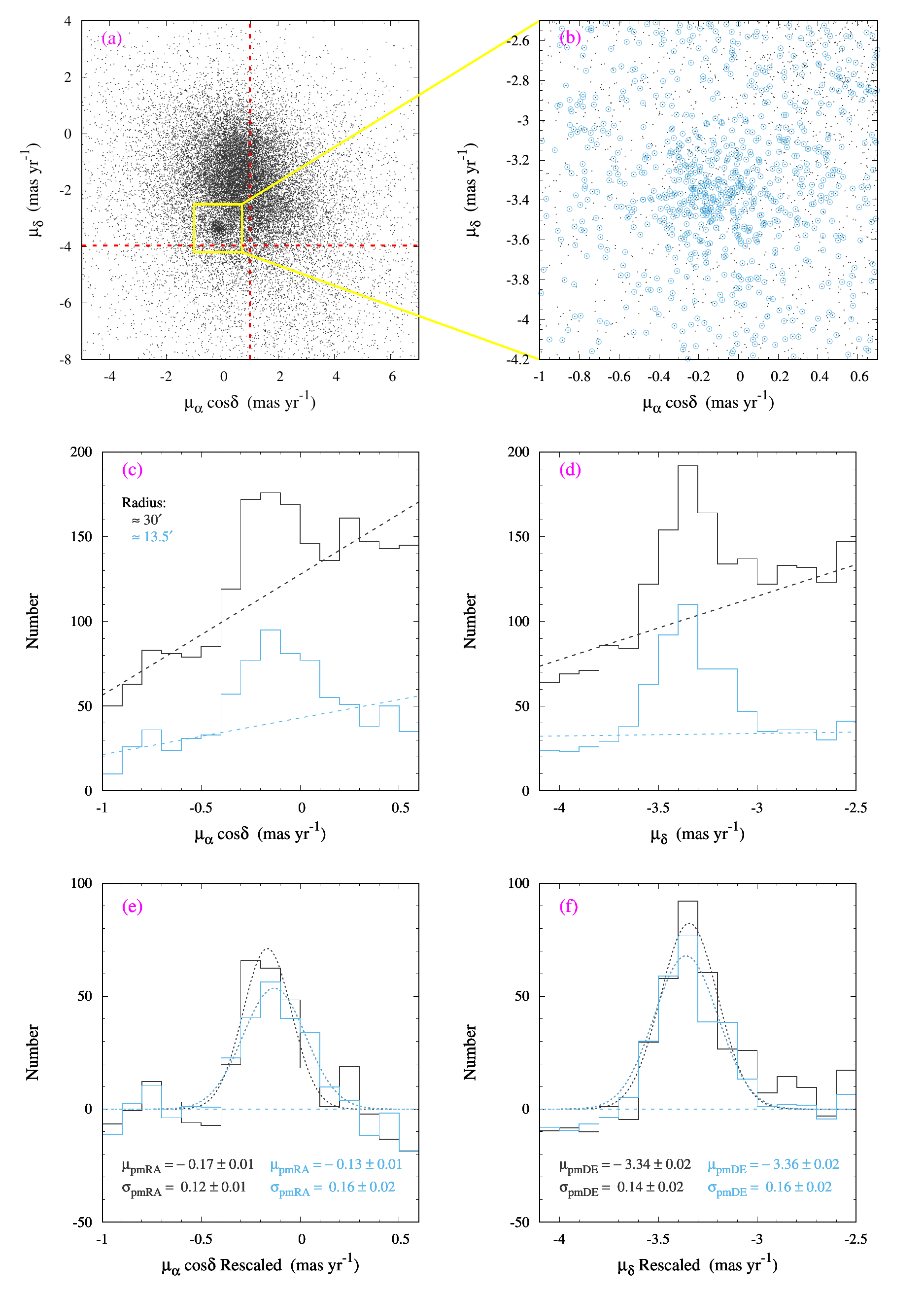}
  \caption{(a): The proper motion vector plot for all the sources toward M\,36 within a radius of $30\arcmin$. Proper motion of M\,36 as listed in SIMBAD \citep{lok03} is presented by red dashed lines, which deviates from our analysis based on {\it Gaia} DR2 results.  (b): Enlargement to show the yellow square in (a), that is, a width of 1.5~mas~yr$^{-1}$ centered around the most concentrated range. (c): Histograms of $\mu_{\alpha} \cos\delta$ in (b) contrasting the enhancement in the cluster region and in the field.  (d): The same as in (c) but for $\mu_{\delta}$.   (e): The background subtracted rescaled distributions, together with a Gaussian fit, of (c). The black and blue colors represent the sources within $30\arcmin$ and $13\farcm$5, respectively. (f): The same as in (e) but for the distributions in (d).  
  See text (Section~\ref{ssec:ast_mem}) for details.}
  \label{fig:pm_dist}
\end{figure*}

The velocity dispersion of the cluster members is typically much lower than that of the field stars, so the members can be generally distinguished by their uniform kinematics \citep{kra07}. The kinematic membership is tested on the basis of the {\it Gaia} DR2 proper motion and parallax data. The proper motion distribution of all the sources within a radius of $30\arcmin$ from the center ($\alpha_{2000}$ = $05^{\rm{h}}36^{\rm{m}}18^{\rm{s}}$, $\delta_{2000}$ = $+ 34^{\rm{d}}08^{\rm{m}}24^{\rm{s}}$) of M\,36 is shown in Figure~\ref{fig:pm_dist}(a). There are two distinct enhancements, one for the cluster members ($\mu_{\alpha} \cos\delta \approx -0.2$~mas~yr$^{-1}$, $\mu_{\delta} \approx -3.4$~mas~yr$^{-1}$), and the other for field stars ($\mu_{\alpha} \cos\delta \approx 0.48$~mas~yr$^{-1}$, $\mu_{\delta} \approx -1.76$~mas~yr$^{-1}$). Conceivably, around the peak for the cluster, the distribution is dominated by members, whereas away from peak the contamination by field stars becomes prominent. A clear separation between the cluster and field motion enables us to select member stars of M\,36 relatively reliably. 
 
The proper motion distribution around the cluster peak is analysed for a square region of width 1.5~mas~yr$^{-1}$. The zoomed-in distributions for all the sources within this box region for a radius of $30\arcmin$ and of $13\farcm5$ from the cluster center are shown in Figure~\ref{fig:pm_dist}(b).  We estimated the relative contribution of members versus field stars by projecting the distribution along $\mu_{\alpha}~cos\delta$ and $\mu_{\delta}$. Each background subtracted rescaled histogram was then fitted with a Gaussian function to quantify the proper motion centroid.  
The expectation values ($\mu$) of the Gaussian fits, assuming a radius of $30\arcmin$,  give $\mu_{\alpha} \cos\delta = -0.17 \pm 0.01$~mas~yr$^{-1}$ and $\mu_{\delta} = -3.34 \pm 0.02$~mas~yr$^{-1}$ with standard deviations ($\sigma$) of $0.12 \pm 0.01$ and $0.14 \pm 0.02$~mas~yr$^{-1}$, respectively.  Likewise, should a radius of $13.5\arcmin$ be adopted, we derived $\mu_{\alpha}~\cos\delta = -0.13 \pm 0.01$~mas~yr$^{-1}$ and $\mu_{\delta} = -3.36 \pm 0.02$~mas~yr$^{-1}$ with standard deviations of $0.16 \pm 0.02$ and $0.16 \pm 0.02$~mas~yr$^{-1}$, respectively.
We adopted the proper motion center of the cluster as the mean of the above two values, i.e., $\mu_{\alpha} \cos\delta = -0.15 \pm 0.01$~mas~yr$^{-1}$ and $\mu_{\delta} = -3.35 \pm 0.02$~mas~yr$^{-1}$. The optimal range of proper motions is selected within a radius of 0.5~mas~yr$^{-1}$ ($\simeq 3\sigma$) from this center.  
In Figure~\ref{fig:pm_dist}(e) and \ref{fig:pm_dist}(f), the normalization of the histograms permits us to estimate the integrated number of members along $\mu_{\alpha}~cos\delta$ as 208, and along $\mu_{\delta}$ as 257, enabling an estimate of the number of missing members (incompleteness) after the parallax constraint is introduced.

\begin{figure*}
\centering
\includegraphics[width=0.7\textwidth]{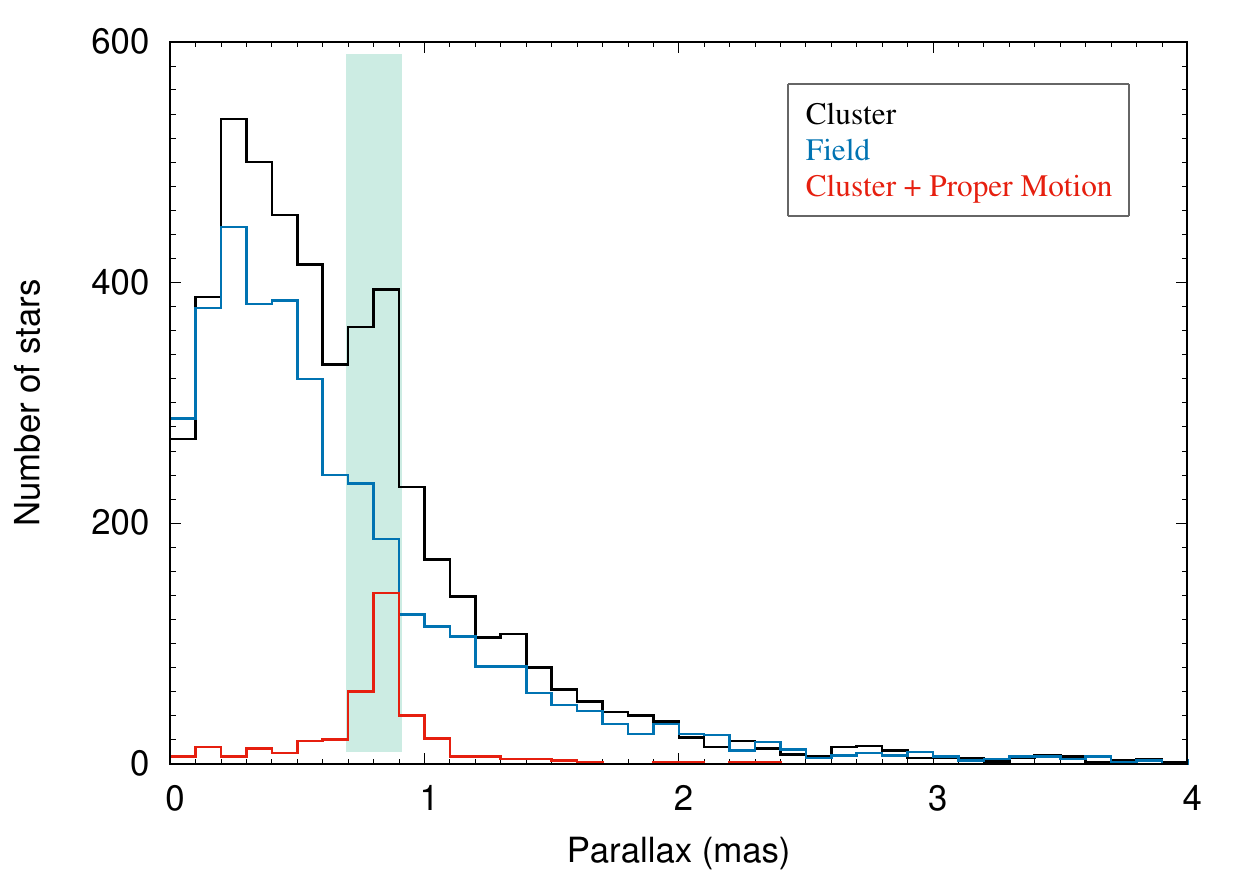}
  \caption{Parallax for all the sources, marked by the histogram in black, within the cluster region (radius of $13\farcm5$), and for those within the control field with the same sky area as the cluster region, marked in blue. The sources within the cluster region and also satisfy the proper motion criteria are plotted in red.  
  The shade depicts the parallax range adopted for the cluster. }
   \label{fig:plx_dis}
\end{figure*}

We further constrained the membership by including {\it Gaia} DR2 parallax measurements. 
Figure~\ref{fig:plx_dis} compares the parallaxes for stars in the cluster region (radius $\simeq 13\farcm5$), with those in a nearby control field, centered around R.A.~$= 05^{\rm h}36^{\rm m}14^{\rm s}$, Decl.~$= +33\degr21\arcmin22\arcsec$ (J2000) ($\ell = 175\fdg1877$; $b = 0\fdg6383$), roughly $47\arcmin$ south, with the same sky area as the cluster region.  Overplotted are the stars that are located within the cluster region, and satisfying the proper motion criteria (average proper motion $< 0.5$~mas~yr$^{-1}$ from the proper motion center).  
It is assuring to see a sharp rise in the parallax range between $0.70 \pm 0.11$~mas and $0.90 \pm 0.08$~mas.  
The parallax range 0.7--0.9~mas is thus adopted to further refine the membership selection.  
The number of stars satisfying the positional (inside the cluster region), kinematic (inside the proper motion range), and parallax criteria are 207, which follows to a variation in the incompleteness factor of $\sim$1--19\%.

\subsection{Cluster Members}
 \label{ssec:clu_mem}

\begin{figure*}
\centering
  \includegraphics[width=0.7\textwidth]{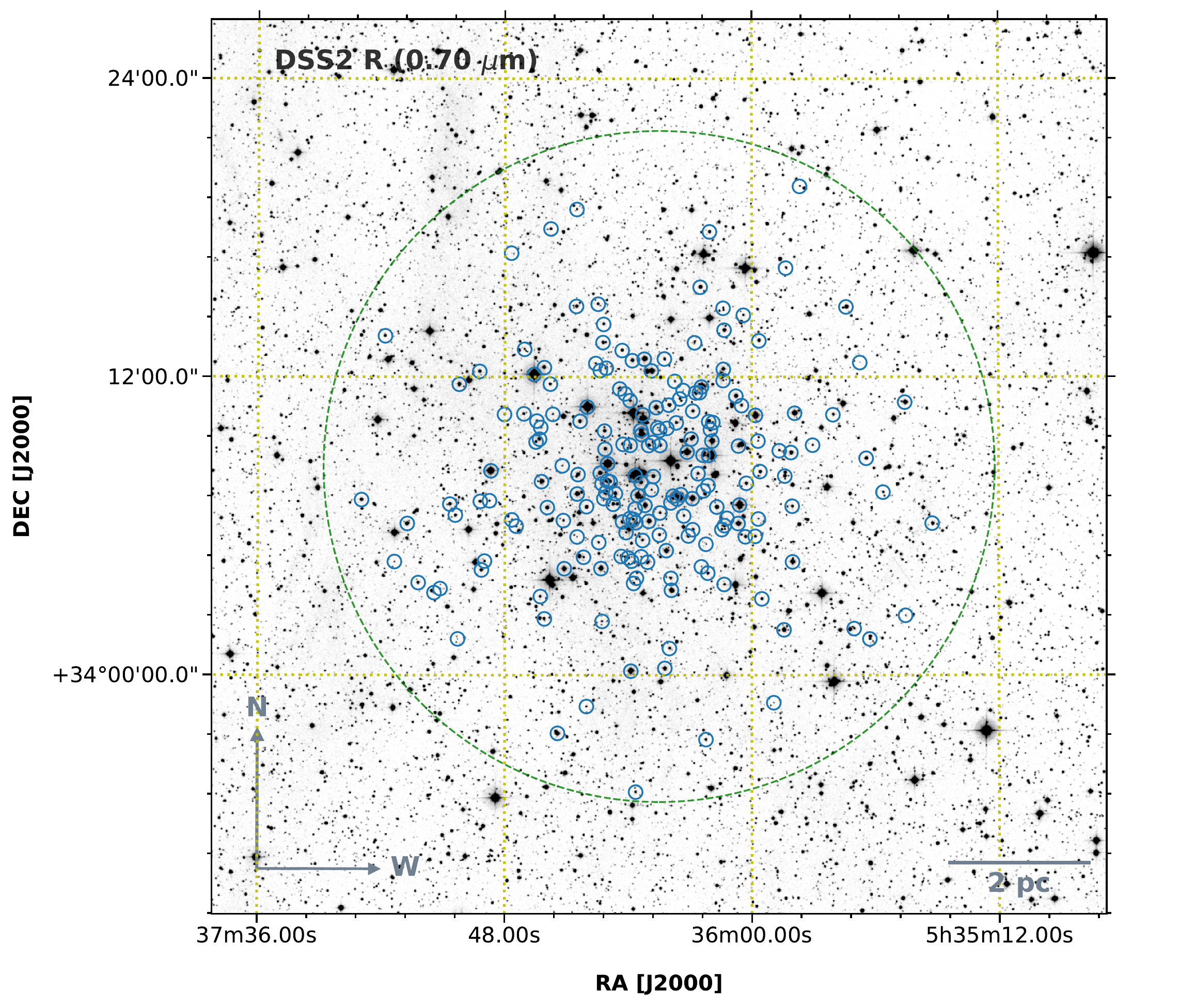}
  \caption{Spatial distribution of the member candidates of M\,36 (marked with blue circles), overlaid on the DSS2~$R$~0.70~$\mu$m image. The green dashed circle represents the cluster extent of a radius of $13\farcm5$.}
  \label{fig:members_spatial}
\end{figure*}

We have identified a total of 200 member candidates by imposing the astrometric and kinematic criteria within the cluster region, that is, within a radius of $\simeq 13\farcm5$, with proper motion less than  0.5~mas~yr$^{-1}$ from the cluster proper motion center ($\mu_{\alpha}$ $cos\delta = -0.15$~mas~yr$^{-1}$, $\mu_{\delta} = -3.35$~mas~yr$^{-1}$), and parallax between 0.7~mas to 0.9~mas.  The median of the parallax for the members is obtained as $0.82 \pm 0.07$~mas, corresponding to a distance of $1.20 \pm 0.13$~kpc.  The photometric and astrometric parameters of all the member candidates are listed in Table~\ref{tab:members}. The spatial distributions of the members are shown in Figure~\ref{fig:members_spatial}.
Among a total of 7948 sources within the cluster region, 757 are found to satisfy the parallax criteria alone and 402 sources satisfy the proper motion criteria alone, with 200 sources satisfying all the above criteria.  It is to be noted that, among the members list, 94\% sources have parallax errors less than 20\%. Moreover, for those sources that satisfy the membership criteria, but have higher parallax errors (20\% or above), we compared their distribution with the theoretical isochrones in both color-magnitude diagrams (Figure~\ref{fig:cmd_mul}(b) and \ref{fig:cmd_mul}(e)), to be presented in the next Section. Additionally 12 sources are thus included as members.  In total, among the 200 members, 182 sources have reliable photometry and 18 sources have higher photometric uncertainty in any of the $J$, $H$, $K$, $[3.6]$, or $[4.5]$ bands.

Our selection does not completely reject field stars; some fraction of the candidates may well be field stars.  To estimate the level of contamination, we choose the same control field as in the previous section, with the same sky area, and exercised the same set of selection criteria as in the cluster region. We  obtained 16 sources in the control field that share similar kinematics. This gives a false positive rate of 8\% in the candidate sample, indicating statistically 184 of the 200 sources to be true members of the cluster.

\newpage
\begin{longrotatetable}

\begin{deluxetable*}{cccccccccccccccc}
\tablecaption{Photometry and astrometry of members}
\tabletypesize{\scriptsize}
\label{tab:members}
\tablehead{
\colhead{ID} & \colhead{R.A. (J2000)} & \colhead{Decl. (J2000)} & 
 \colhead{$J$} & \colhead{$H$} & \colhead{$K$} & \colhead{[3.6]} & \colhead{[4.5]} & 
 \colhead{$G$} & \colhead{$G_{BP}$} & \colhead{ $G_{RP}$} & 
 \colhead{$\mu_{\alpha} \cos\delta$}  & \colhead{ $\mu_{\delta}$}  & \colhead{Parallax} &  \colhead{Distance} & \colhead{RV}  \\
  \colhead{No.} & \colhead{(deg)} & \colhead{(deg)} & \colhead{(mag)} & 
  \colhead{(mag)} & \colhead{(mag)} & \colhead{(mag)} & \colhead{(mag)} & 
  \colhead{(mag)} & \colhead{(mag)} & \colhead{(mag)} & 
  \colhead{(mas~yr$^{-1}$)} & \colhead{(mas~yr$^{-1}$)} & \colhead{(mas)} &  \colhead{(kpc)} & \colhead{(km~s$^{-1}$)} 
          }
\startdata
  1  &  84.039139  &  34.123608  &  13.622  &  13.238  &  13.117  &  13.050  &  12.993  &  15.048  &  15.522  &  14.399  &  -0.113  &  -3.539  &  0.8658  &  1.120  &  \nodata  \\
  &  &  & $\pm$ 0.026  & $\pm$ 0.028  & $\pm$ 0.030  & $\pm$ 0.038  & $\pm$ 0.036  & $\pm$ 0.002  & $\pm$ 0.007  & $\pm$ 0.005  & $\pm$ 0.072  & $\pm$ 0.052  & $\pm$ 0.0369  & $ _{-0.04}^{+0.05}$  &  \nodata  \\
  2  &  84.141708  &  34.312340  &  14.709  &  14.195  &  14.042  &  13.786  &  13.798  &  16.460  &  17.091  &  15.688  &  0.033  &  -3.630  &  0.8516  &  1.146  &  \nodata  \\
  &  &  & $\pm$ 0.003  & $\pm$ 0.003  & $\pm$ 0.005  & $\pm$ 0.049  & $\pm$ 0.059  & $\pm$ 0.004  & $\pm$ 0.025  & $\pm$ 0.007  & $\pm$ 0.158  & $\pm$ 0.111  & $\pm$ 0.0811  & $ _{-0.10}^{+0.12}$  &  \nodata  \\
  3  &  84.270081  &  34.062000  &  13.341  &  13.043  &  12.932  &  12.824  &  12.823  &  14.546  &  14.917  &  14.003  &  -0.086  &  -3.008  &  0.7292  &  1.322  &  \nodata  \\
  &  &  & $\pm$ 0.021  & $\pm$ 0.022  & $\pm$ 0.026  & $\pm$ 0.035  & $\pm$ 0.037  & $\pm$ 0.000  & $\pm$ 0.002  & $\pm$ 0.001  & $\pm$ 0.061  & $\pm$ 0.045  & $\pm$ 0.0357  & $ _{-0.06}^{+0.07}$  &  \nodata  \\
  4  &  84.220413  &  34.203709  &  13.458  &  13.065  &  12.957  &  12.858  &  12.822  &  14.829  &  15.260  &  14.229  &  -0.045  &  -3.414  &  0.8494  &  1.141  &  \nodata  \\
  &  &  & $\pm$ 0.023  & $\pm$ 0.022  & $\pm$ 0.024  & $\pm$ 0.043  & $\pm$ 0.047  & $\pm$ 0.001  & $\pm$ 0.003  & $\pm$ 0.002  & $\pm$ 0.063  & $\pm$ 0.048  & $\pm$ 0.0362  & $ _{-0.05}^{+0.05}$  &  \nodata  \\
  5  &  84.065506  &  34.064941  &  13.849  &  13.517  &  13.415  &  13.292  &  13.303  &  15.273  &  15.735  &  14.645  &  -0.124  &  -3.478  &  0.8375  &  1.158  &  \nodata  \\
  &  &  & $\pm$ 0.001  & $\pm$ 0.001  & $\pm$ 0.003  & $\pm$ 0.043  & $\pm$ 0.041  & $\pm$ 0.001  & $\pm$ 0.005  & $\pm$ 0.004  & $\pm$ 0.078  & $\pm$ 0.055  & $\pm$ 0.0505  & $ _{-0.07}^{+0.07}$  &  \nodata  \\
  6  &  84.093369  &  34.065090  &  13.426  &  13.166  &  13.041  &  12.982  &  12.952  &  14.611  &  14.990  &  14.065  &  -0.010  &  -3.149  &  0.8782  &  1.104  &  -19.89  \\
  &  &  & $\pm$ 0.023  & $\pm$ 0.026  & $\pm$ 0.026  & $\pm$ 0.036  & $\pm$ 0.029  & $\pm$ 0.000  & $\pm$ 0.002  & $\pm$ 0.002  & $\pm$ 0.056  & $\pm$ 0.040  & $\pm$ 0.0313  & $ _{-0.04}^{+0.04}$  &  8.80  \\
  7  &  84.094398  &  34.133839  &  9.035  &  9.081  &  9.088  &  9.135  &  9.064  &  9.078  &  9.096  &  9.044  &  -0.151  &  -3.451  &  0.7819  &  1.241  &  \nodata  \\
  &  &  & $\pm$ 0.022  & $\pm$ 0.019  & $\pm$ 0.018  & $\pm$ 0.039  & $\pm$ 0.034  & $\pm$ 0.001  & $\pm$ 0.002  & $\pm$ 0.002  & $\pm$ 0.131  & $\pm$ 0.097  & $\pm$ 0.0646  & $ _{-0.09}^{+0.11}$  &  \nodata  \\
  8  &  84.114609  &  34.129707  &  13.381  &  13.129  &  13.073  &  13.017  &  12.899  &  14.607  &  14.976  &  14.053  &  -0.371  &  -3.372  &  0.8466  &  1.144  &  \nodata  \\
  &  &  & $\pm$ 0.029  & $\pm$ 0.030  & $\pm$ 0.029  & $\pm$ 0.040  & $\pm$ 0.040  & $\pm$ 0.000  & $\pm$ 0.002  & $\pm$ 0.002  & $\pm$ 0.053  & $\pm$ 0.039  & $\pm$ 0.0300  & $ _{-0.04}^{+0.04}$  &  \nodata  \\
  9  &  84.121284  &  34.128357  &  12.934  &  12.707  &  12.614  &  12.646  &  12.505  &  13.989  &  14.300  &  13.512  &  -0.002  &  -3.314  &  0.8558  &  1.133  &  \nodata  \\
  &  &  & $\pm$ 0.022  & $\pm$ 0.022  & $\pm$ 0.023  & $\pm$ 0.039  & $\pm$ 0.033  & $\pm$ 0.000  & $\pm$ 0.002  & $\pm$ 0.001  & $\pm$ 0.057  & $\pm$ 0.041  & $\pm$ 0.0383  & $ _{-0.05}^{+0.05}$  &  \nodata  \\
  10  &  84.211266  &  34.137028  &  \nodata  &  11.372  &  9.986  &  9.923  &  9.877  &  10.188  &  10.251  &  10.071  &  -0.588  &  -3.433  &  0.8461  &  1.147  &  \nodata  \\
  &  &  &  \nodata  & $\pm$ 0.000  & $\pm$ 0.000  & $\pm$ 0.038  & $\pm$ 0.023  & $\pm$ 0.001  & $\pm$ 0.002  & $\pm$ 0.003  & $\pm$ 0.096  & $\pm$ 0.073  & $\pm$ 0.0485  & $ _{-0.06}^{+0.07}$  &  \nodata  \\  
\enddata
\end{deluxetable*}

\end{longrotatetable}

\subsection{Color-Magnitude Diagram Analysis}
 \label{ssec:cmd_ana}

\begin{figure*}
\centering
  \includegraphics[width=\textwidth]{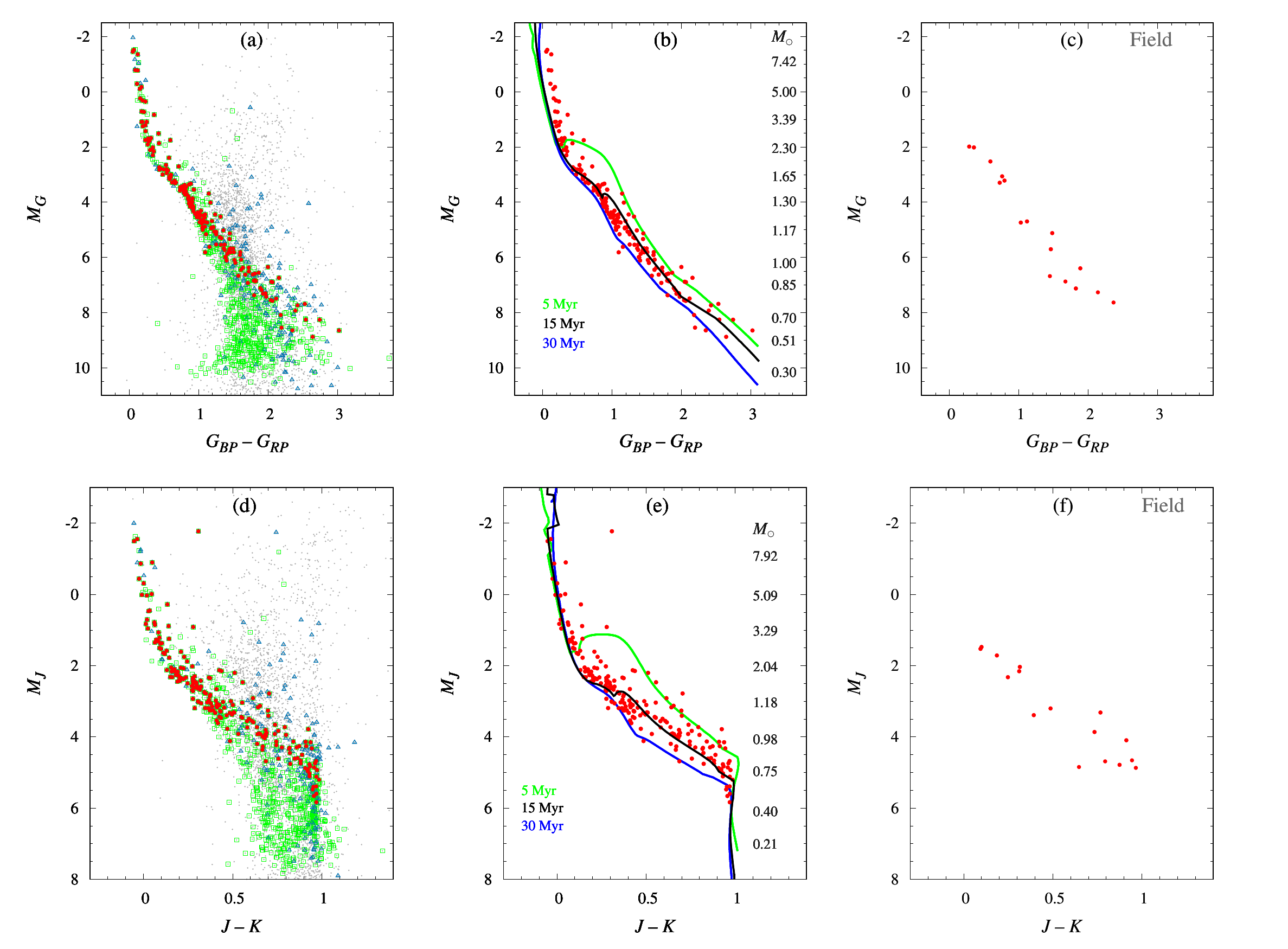}
  \caption{Color-magnitude diagrams for (top panels) the {\it Gaia}~DR2 data, and (bottom panels) the UKIDSS data. The gray dots represent all in the cluster region, whereas the blue symbols mark the stars passing the proper motion selection only, the green ones are stars passing the parallax selection only, and the red symbols are stars satisfy both criteria, so are member candidates. The central panels compares the member candidates with the PARSEC isochrones of ages 5, 15, and 30~Myr, assuming a reddening correction $E(B-V) = 0.25$~mag \citep{san00} and metallicity [Fe/H]$=-0.15$. The stellar masses according to the 15~Myr isochrone are indicated.}
  \label{fig:cmd_mul}
\end{figure*}

Figure~\ref{fig:cmd_mul} compares the color-magnitude diagram for stars seen toward the cluster region with that toward the control field region, with photometric data collected from {\it Gaia}~DR2 and UKIDSS, and converted to absolute magnitudes by adopting the distance values  computed by \citet{bai18} from the {\it Gaia} parallax, which provides empirically better estimate of distances for all the stars with parallaxes published in the {\it Gaia}~DR2, using a probabilistic inference approach, which takes into account for the nonlinearity of the transformation and the positivity constraint of distance.  

The average age of the members is estimated by comparing with theoretical PARSEC isochrones \citep{bre12, mar17, pas19}.  The photometric sensitivity curves from \citet{mai18} are assumed for the {\it Gaia} sources, as they give empirically the best fit to our data.  The majority of the members are consistent with an age between 5 to 30 Myr, with a best fit age of 15 Myr as seen in Figures~\ref{fig:cmd_mul}(b) and \ref{fig:cmd_mul}(e).  The estimated age is slightly younger than the values reported in literature (Section~\ref{sec:intro}), where the age consistently varied from 20 Myr to 30 Myr. The candidate sample in this analysis is limited down to $G \sim 20.3$~mag, $G_{BP} \sim 21.1$~mag, $G_{RP} \sim 19.0$~mag, $J \sim 17.6$~mag, and $K \sim 16.7$~mag, equivalent to member masses $\gtrsim0.6$~M$_\sun$.

\subsection{Luminosity Function and Mass Function}
 \label{ssec:lum_mas}

\begin{figure*}
\centering
  \includegraphics[width=\textwidth]{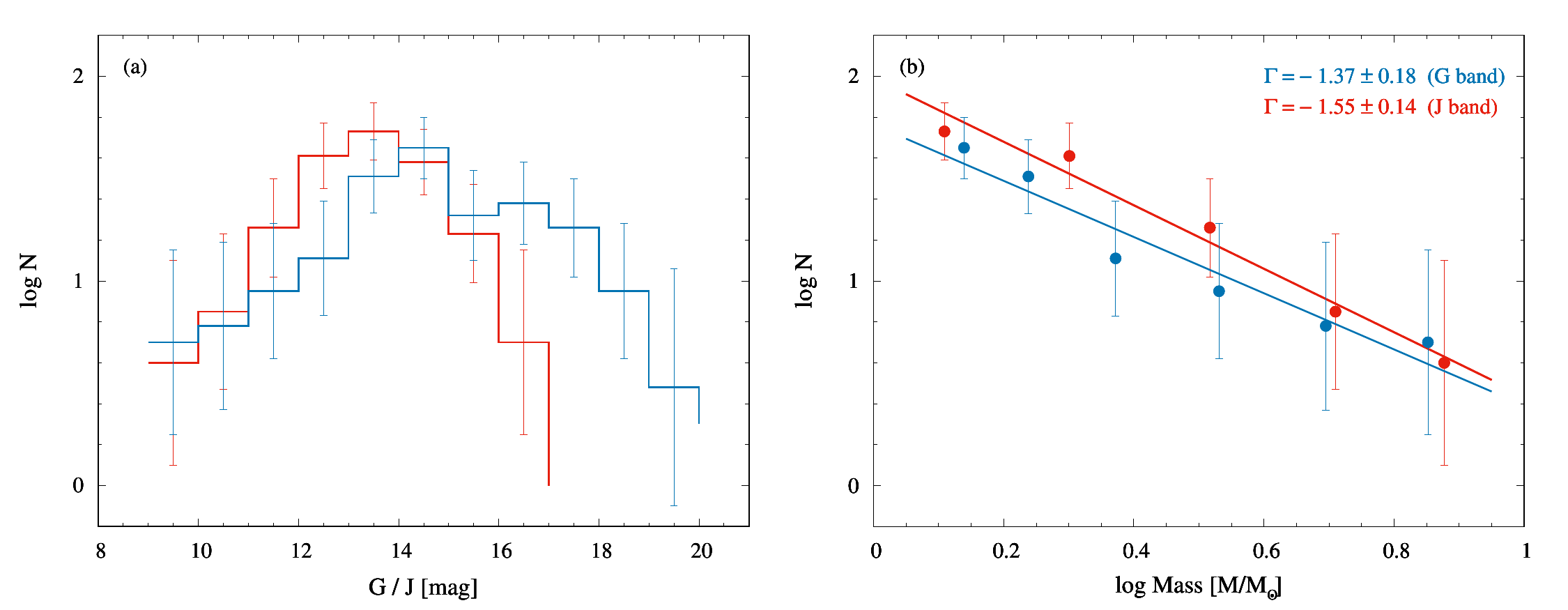}
  \caption{(a)~The luminosity function of the cluster generated with $G$ (blue) or $J$ (red) band, after removing field star contamination. The error bars represent Poisson errors. 
  (b)~The mass function of the cluster derived from the luminosity function using 
  the PARSEC isochrones of an age of 15~Myr, adjusted according to the cluster distance and reddening.}
  \label{fig:lum_mas}
\end{figure*}

The luminosity function of the cluster is derived by statistically subtracting the number of stars in the control field from that toward the cluster region, for each magnitude bin. Field stars were chosen on the basis of satisfying the same proper motion and parallax criteria as for the member candidates. The decontaminated distribution of the luminosity function, generated with both the $G$ and $J$ bands is depicted in Figure~\ref{fig:lum_mas}(a). The luminosity function peaks around 14--15~mag for $G$ band and 13--14 mag for $J$ band, after which it declines swiftly owing to the data incompleteness. 

The mass function of the cluster was derived on the basis of luminosity function and is presented in Figure~\ref{fig:lum_mas}(b). Member masses were estimated according to the PARSEC isochrones of an age of 15~Myr (Section~\ref{ssec:cmd_ana}), corrected for a distance of 1.22~kpc (Section~\ref{ssec:clu_mem}) and reddening $E(B-V) = 0.25$~mag \citep{san00}, with a polynomial interpolation of discrete data points. The slope of a power-law least-squares fit to the mass function, expressed as $\Gamma = d \log N(\log m) / d \log m$, where $N(\log m)$ is the number of stars per unit logarithmic mass, up to the photometric sensitivity, gives a slope of $\Gamma = -1.37 \pm 0.18$ for the $G$ band, and a slightly steeper $\Gamma = -1.55 \pm 0.14$ for the $J$ band. The mass range used to derive the mass function is $1.38 \lesssim M/M_\sun \lesssim 7.11$ for the $G$ band and $1.28 \lesssim M/M_\sun \lesssim 7.54$ for the $J$ band. 

In comparison to literature works, \citet{san00} derived the cluster mass function with a  slope of $\Gamma = -1.23 \pm 0.17$ for the mass interval $0.72 \lesssim M/M_\sun \lesssim 9.40$, on the basis of proper motion membership and statistical field star subtraction. Using the $V$ band photometry and a statistical field star subtraction approach, \citet{sha08} reported a slope of $\Gamma = -1.80 \pm 0.14$ for the mass range $1.01 \lesssim M/M_\sun \lesssim 6.82$. Recently, \citet{jos20} estimated $\Gamma = -1.26 \pm 0.19$ for the mass range $0.72 \lesssim M/M_\sun \lesssim 7.32$, by considering members only.  Our results, based on reliable membership above $\sim1$ solar mass selected with {\it Gaia} parallax and proper motions plus contamination subtraction, fall within the values published in the literature, and are consistent with the canonical value of $\Gamma =-1.35$ in the solar neighborhood \citep{sal55}.

\section{The Young Stellar Population}
 \label{sec:ext_yso}

The overall distribution of dust in a cloud can be traced by measuring the extinction of background starlight \citep{lad94}. However, because the dust clouds are often clumpy, it is challenging to get a complete census of detectable background stars, particularly toward embedded sources \citep{gut05}.  The situation is mitigated in near-infrared wavelengths where the dust opacity is modest and the reddening law appears universal \citep{jon80, mar90, whi93, fla07}. We have used measurements of infrared color excess, together with certain aspects of stellar number counts of background stars, to map the extinction distribution throughout the cloud.

\subsection{Extinction Map}
 \label{ssec:ext_map}

Given that optical extinction decreases with increasing wavelength, observations made at longer wavelengths can detect more background stars through a cloud, and probe deeper cloud depths \citep{str89, dic90}. We utilized the $H$ and $K$ band photometry from the UKIDSS catalog, and constructed a number density image by defining a grid over the target area, following the method outlined in \citet{gut05}.  Briefly, the region was divided into uniform grids of size  $5\arcsec\times5\arcsec$. The 20 nearest-neighbor sources from the center of each grid were selected to calculate the mean and standard deviation of the ($H-K$) color for each grid, excluding the sources for which the ($H-K$) values deviated $\gtrsim 3\sigma$ from the mean value. The mean ($H-K$) color for each grid then was converted to $A_{K}$, using the reddening law $A_{K} = 1.82 \times [(H-K)_{\rm obs}-(H-K)_{\rm int}]$, the difference between the observed and the intrinsic color \citep{fla07}. 

\begin{figure*}
\centering
  \includegraphics[width=\textwidth]{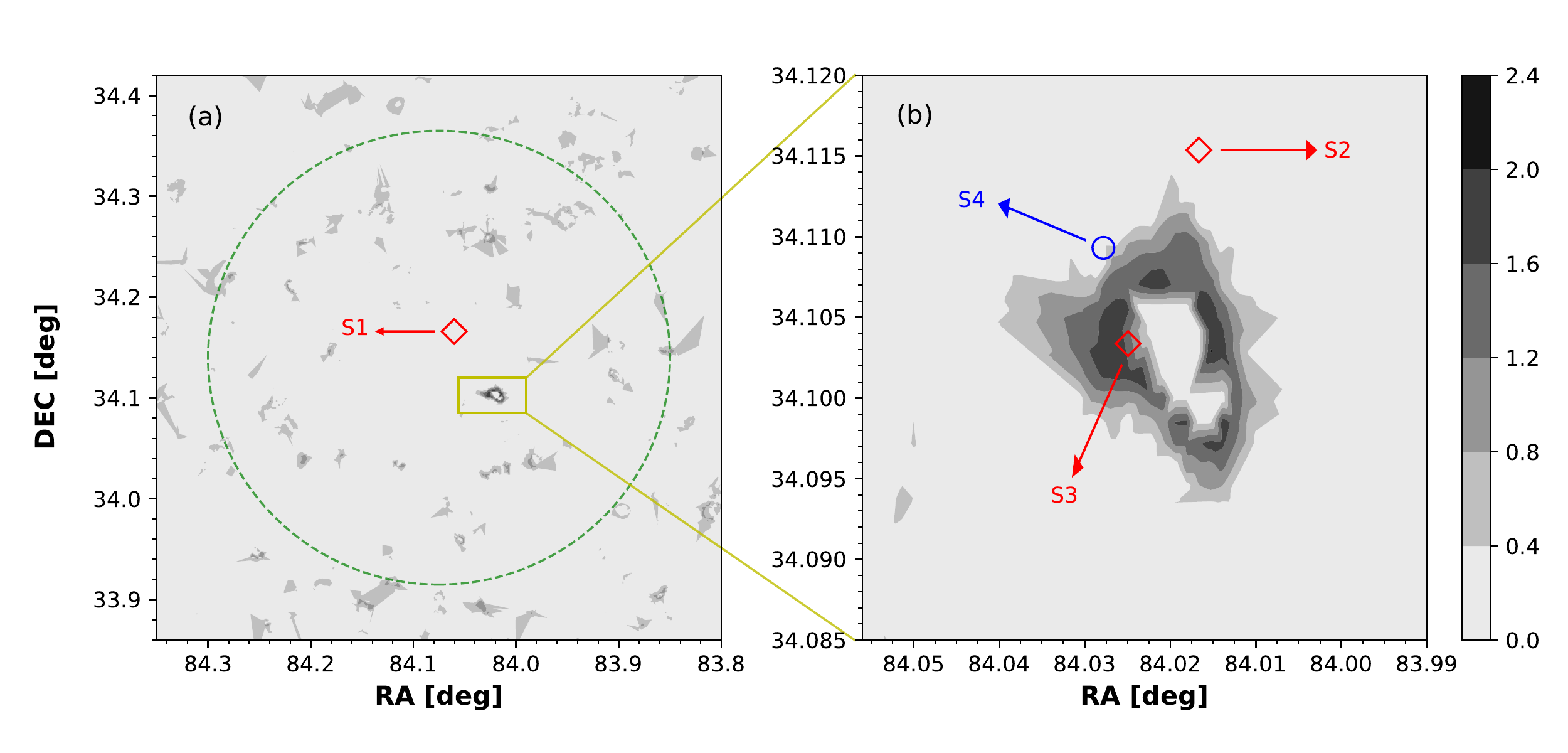}
  \caption{(a) The $K$-band extinction map generated using the UKIDSS photometry. The green dashed circle depicts the $13\farcm5$ cluster radius. The yellow box marks the region of very high extinction, with $A_{V}$ up to $\sim 22.8$~mag. (b) The extinction distribution of the yellow box in (a), in which the S1, S2, and S3 (all Class I objects) and S4 (a Class II object) YSOs are identified  (Section~\ref{ssec:yso_iden}). The $A_K$ values are represented by the gray-scale contours.}
  \label{fig:ext_map}
\end{figure*}

From a nearby comparison field with little extinction, the average intrinsic color $(H-K)_{\rm int}$ was estimated to be $\sim 0.2$~mag. To restrict our analysis as far as possible to background objects, we have selected only sources with little extinction ($J-H<1.0$~mag, $H-K<0.6$~mag) for analysis \citep{pan20}.  The resulting extinction map is displayed in Figure~\ref{fig:ext_map}. The derived extinction values range from $A_{V} \simeq 1.33$--22.82~mag, or $A_{K} \simeq 0.12$--2.05~mag.  
Inspection of the extinction map reveals a compact region ($\sim 1 \farcm9 \times 1\farcm2$, centering around $\alpha_{2000} = 84\fdg023$, $\delta_{2000} = + 34\fdg103$) with excessive extinction ($A_{V} \sim 22.8$~mag).  The rest of the cluster has otherwise relatively low extinction, with $A_V \lesssim 4.56$~mag.

An extinction map thus produced is limited in angular resolution by the availability of detectable background stars.  Moreover, the extinction value along a particular line of sight is estimated in a statistical manner \citep{lad94}. Empirically, we found a $\sim5\arcsec$ grid size, and $\sim20$ nearest neighbor stars to be optimal choices, as a compromise between sensitivity and resolution.  Our extinction map with an angular resolution of $5\arcsec$ and sensitivity down to $A_V\sim 27$~mag serves to guide the identification of  heavily embedded sources such as protostellar objects, as will be discussed in the next section.

\subsection{YSOs from the Infrared Photometry}
 \label{ssec:yso_iden}

A certain combination of infrared colors can be used to distinguish YSOs at different evolutionary stages, as the amount of excessive infrared emission arising from the young circumstellar disks and infalling envelopes diminishes with age.  We use such a color-color diagram to diagnose the possible nature of our targets. Lacking longer wavelength IRAC [5.8] and [8.0] $\mu$m detection, we utilized IRAC [3.6] and [4.5] $\mu$m along with UKIDSS $H$ and $K$ band photometry to characterize the YSOs \citep{dut18, pan20} toward M\,36. 

\begin{figure}
\centering
  \includegraphics[width=11.5cm, height=8.0 cm, bb=50 00 400 250]{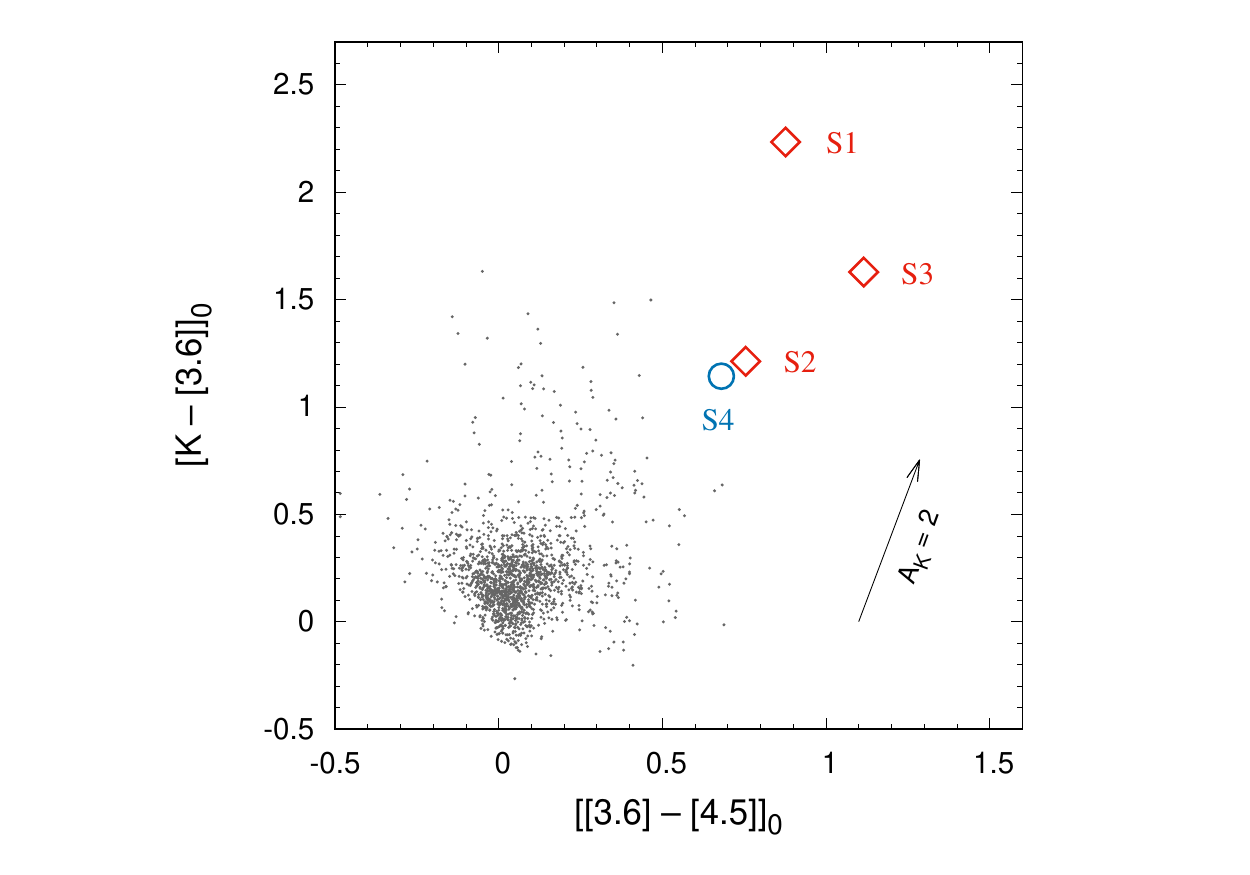}
  \caption{Dereddened color-color diagram from the {\it Spitzer} and UKIDSS photometry, showing the distribution of field stars and main-sequence sources (gray dots), Class II (blue circle), and Class I objects (red diamonds). The black arrow marks the reddening vector for $A_{K} = 2$~mag.  }
  \label{fig:spi_yso}
\end{figure}

It has been shown that [3.6]$-$[4.5] is a useful YSO class discriminant color \citep{all04, meg04}, provided that the colors are dereddened. The dereddened ([[3.6]$-$[4.5]]$_{0}$ versus [$K-$[3.6]]$_{0}$) color-color diagram is displayed in Figure~\ref{fig:spi_yso}. To deredden a source, we estimated its line-of-sight extinction from the extinction map (Section~\ref{ssec:ext_map}) and exercised the reddening laws from \citet{fla07} to compute the dereddened colors. We used the set of equations developed by \citet{gut09} to identify the YSOs based on different color criteria.  To minimise contamination from extragalactic sources, an additional brightness cut was applied on the dereddened [3.6] $\mu$m magnitude by requiring a Class~II object to have [3.6]$_{0}< 14.5$~mag and a Class I object to satisfy [3.6]$_{0} < 15$~mag \citep{gut09}. While the majority of the field stars and main-sequence stars have dereddened colors about 0.0~mag to 0.2~mag, three Class I and one Class II candidates stand out, with three of them in close physical association with, signifying ongoing star formation in, the dusty clump depicted in Figure~\ref{fig:ext_map}.  The photometric parameters of these YSOs are listed in Table~\ref{tab:ysos_phot}.

\begin{longrotatetable}
\begin{deluxetable*}{cccccccccc}
\tablecaption{Photometry and astrometry of YSOs}
\tabletypesize{\scriptsize}
\label{tab:ysos_phot}
\tablehead{
 \colhead{Star} & \colhead{R.A. (J2000)} & \colhead{Decl. (J2000)} & 
 \colhead{$J$} & \colhead{$H$} & \colhead{$K$ }& \colhead{[3.6]} & \colhead{[4.5]} 
   & \colhead{Distance}  & \colhead{Type} \\
 \colhead{ID} & \colhead{(deg)} & \colhead{(deg)} & \colhead{(mag)} & 
 \colhead{(mag)} & \colhead{(mag)} & \colhead{(mag)} & \colhead{(mag)} & \colhead{(kpc)} & 
}
\startdata
S1 & 84.060254 & +34.166009 & 18.922 $\pm$ 0.062 & 17.953 $\pm$ 0.059 & 16.718 $\pm$ 0.040 & 14.435 $\pm$ 0.048 & 13.546 $\pm$ 0.037 & $...$ & Class I \\
S2 & 84.016659 & +34.115367 & 13.73 $\pm$ 0.026 & 12.555 $\pm$ 0.023 & 11.717 $\pm$ 0.020 & 10.441 $\pm$ 0.039 & 9.671 $\pm$ 0.031 & $1.395_{-0.26}^{+0.40}$ & Class I \\
S3 & 84.024939 & +34.103370 & 13.397 $\pm$ 0.033 & 11.975 $\pm$ 0.028 & 10.487 $\pm$ 0.022 & 8.81 $\pm$ 0.041 & 7.683 $\pm$ 0.032 & $...$ & Class I \\
S4 & 84.027821 & +34.109306 & 16.577 $\pm$ 0.008 & 15.357 $\pm$ 0.006 & 14.397 $\pm$ 0.005 & 13.205 $\pm$ 0.045 & 12.513 $\pm$ 0.04 & $...$ & Class II \\
\enddata
\end{deluxetable*}
\end{longrotatetable}

\subsection{Spectral Energy Distribution of the YSOs}
 \label{ssec:sed_yso}

To reveal the underlying stellar and circumstellar properties of the young objects, we fitted their spectral energy distributions (SEDs) using theoretical models \citep{zhx15}. We used the radiative transfer models of \citet{rob06,rob07} to derive the physical parameters of the YSOs. The grids of models cover a wide range of parameters, consisting of pre-main-sequence stars plus a combination of axisymmetric circumstellar disks, infalling flattened envelopes, and outflow cavities. These models also span a large range of possible evolutionary stages (from deeply embedded protostars to stars surrounded only by optically thin disks) and stellar masses (from 0.1 to 50~$M_\sun$). The radiative transfer solution is subject to parameter degeneracy, but fluxes covering a sufficiently wide range of wavelengths can reduce this degeneracy.  

\begin{figure*}
\centering
  \includegraphics[width=0.45\textwidth,angle=0]{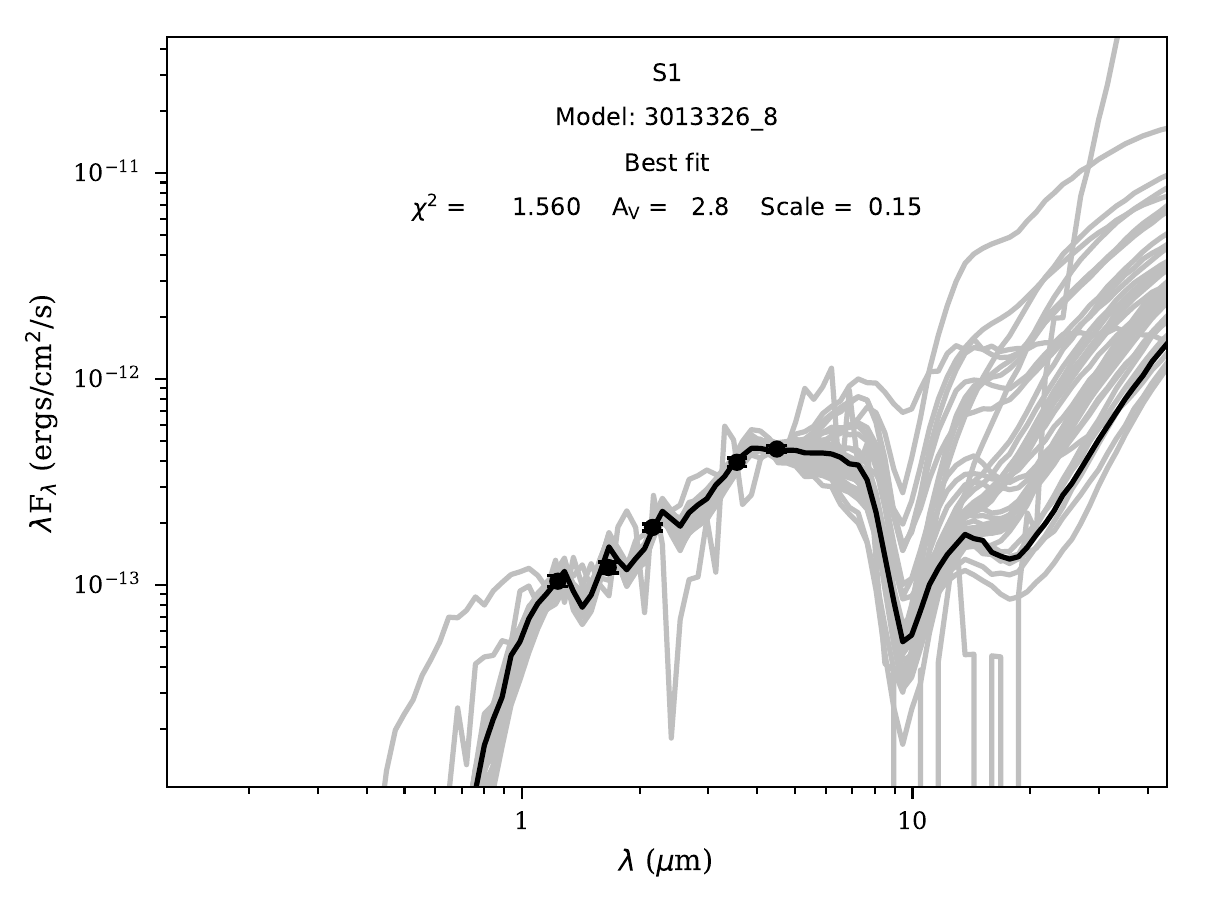}
  \includegraphics[width=0.45\textwidth,angle=0]{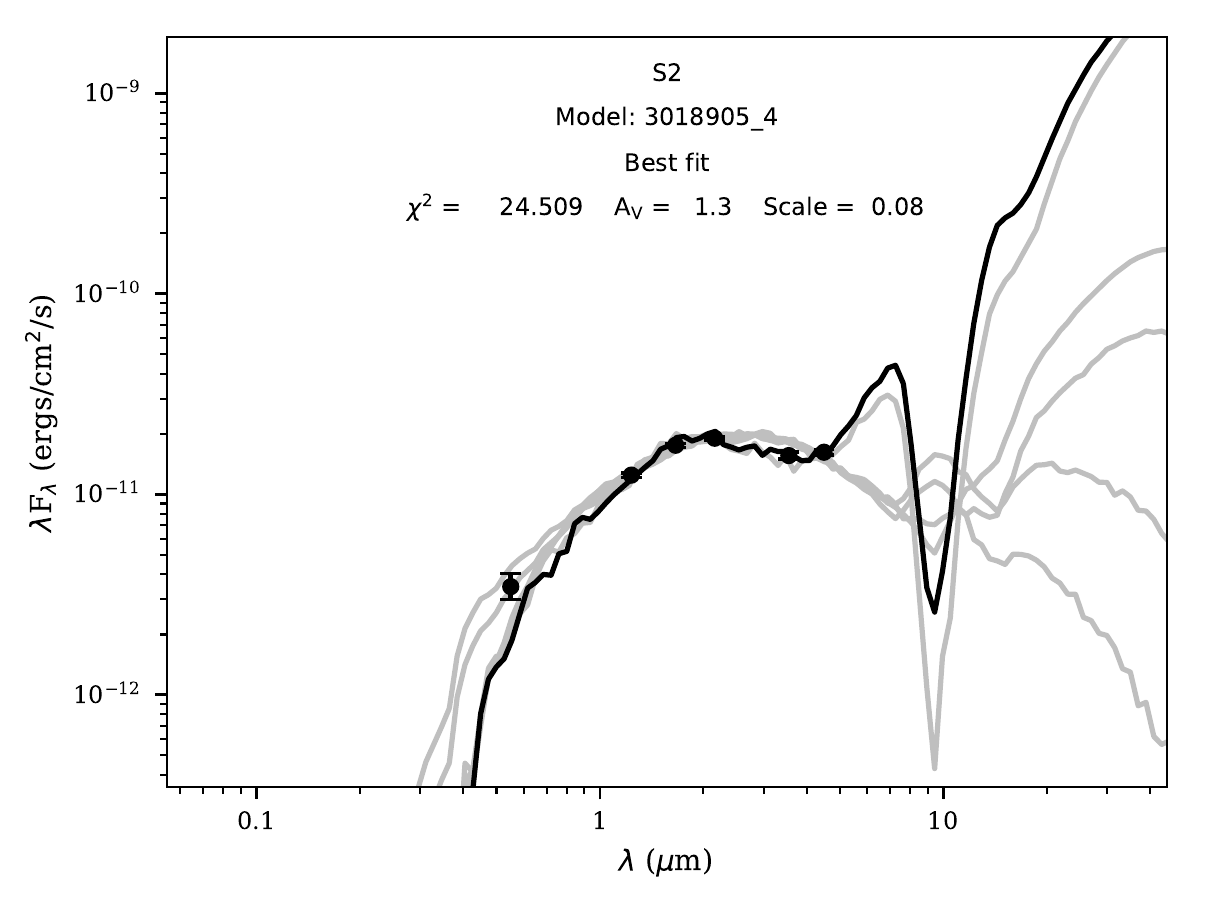}
  \includegraphics[width=0.45\textwidth,angle=0]{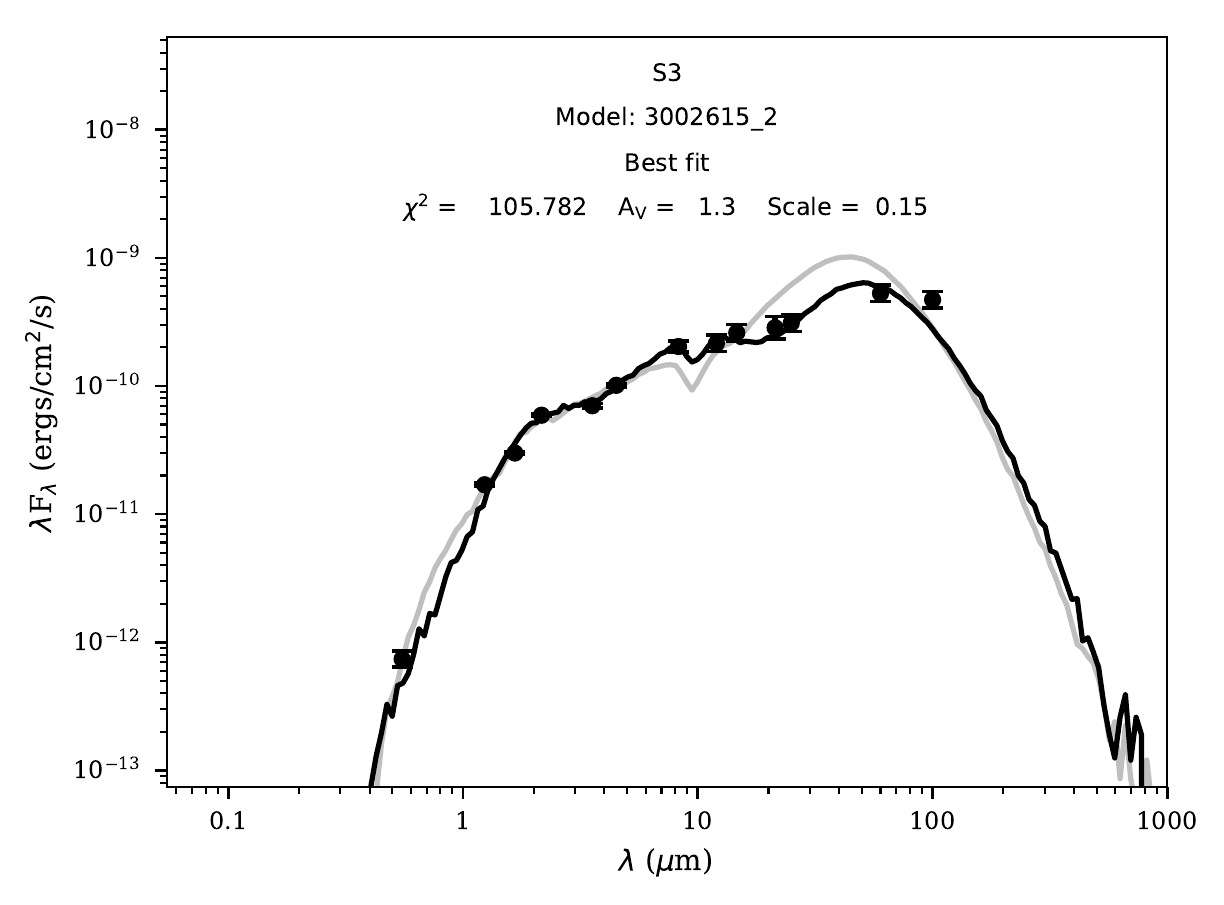}
  \includegraphics[width=0.45\textwidth,angle=0]{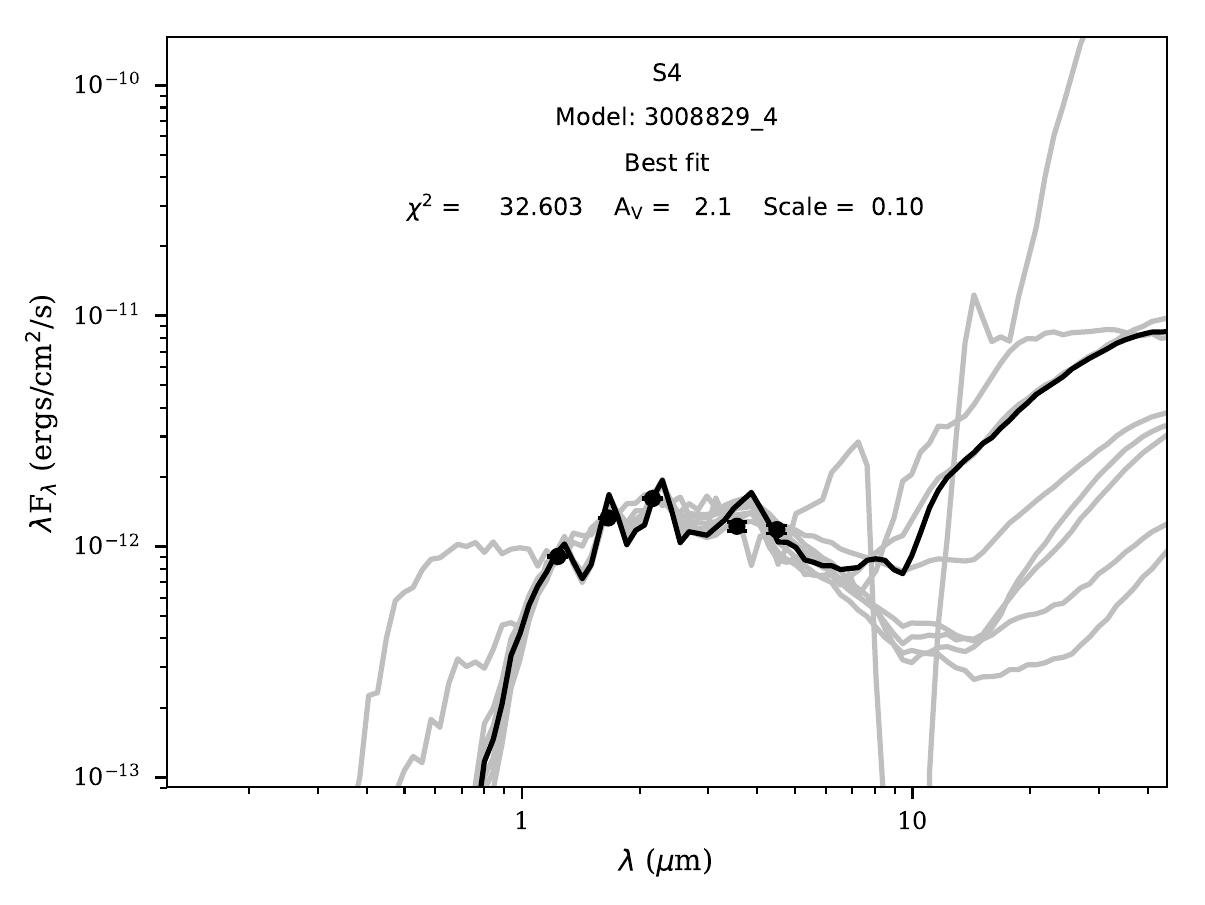}
  \caption{Spectral energy distributions of the YSOs, fitted with the grid models of \citet{rob07}. In each case, the black filled circles are the observed fluxes listed in Table~\ref{tab:flux_ysos}. The solid black and grey lines indicate, respectively, the best-fit model and consecutive good fits for, per data point, $\chi^{2}-\chi^{2}_{\rm best} < 3$.}
  \label{fig:sed_ysos}
\end{figure*}

The photometric fluxes of the YSOs are collected from NOMAD ($V$; \citealt{zac04}), UKIDSS ($J$, $H$, and $K$; \citealt{law07}), IRAC ([3.6] and [4.5]~$\mu$m; \citealt{faz04}), {\it MSX} (8.28, 14.65, and 21.34~$\mu$m; \citealt{sjo09}), and {\it IRAS} (12, 25, 60, and 100~$\mu$m; \citealt{neu84}), wherever available, detailed in Table~\ref{tab:flux_ysos}. We required at least a minimum of five data points for each source to construct the SEDs, which then are fitted with the distance and visual extinction as input parameters to constrain the SEDs. We varied the distance as $1.20 \pm 0.13$~kpc (Section~\ref{ssec:ast_mem}) and extinction $A_V \sim 1.3$~mag to 23~mag (Section~\ref{ssec:ext_map}). 
For each source, the ``best-fit'' model is defined by constraining $\chi^{2}-\chi^{2}_{\rm best}< 3$ (per data point), where $\chi^{2}_{\rm best}$ quantifies the goodness-of-fit parameter. The SEDs and their best-fit models of the four YSOs are shown in Figure~\ref{fig:sed_ysos}.  Expecting that the fitted SEDs are moderately degenerate, we adopted the weighted mean and standard deviations of each physical parameter, and they are given in Table~\ref{tab:sed_par}. The degeneracies obtained for sources S1, S2, S3, and S4 are 36, 6, 2, and 9, respectively.

\begin{longrotatetable}
\begin{deluxetable*}{cccccccccccccc}
\tablecaption{Flux (mJy) from NOMAD ($V$), UKIDSS ($J$, $H$, $K$), IRAC ([3.6]~$\mu$m, [4.5]~$\mu$m), {\it MSX} (8.28~$\mu$m, 14.65~$\mu$m, 21.34~$\mu$m), and {\it IRAS} (12~$\mu$m, 25~$\mu$m, 60~$\mu$m, 100~$\mu$m) catalog.}
\tabletypesize{\tiny}
\label{tab:flux_ysos}
\tablehead{
\colhead{Source} & \colhead{$V$} & \colhead{$J$} & \colhead{$H$} & \colhead{$K$} & \colhead{[3.6]~$\mu$m} & \colhead{[4.5]~$\mu$m} & \colhead{8.28 $\mu$m} & \colhead{14.65 $\mu$m} & \colhead{21.34 $\mu$m} & \colhead{12 $\mu$m} & \colhead{25 $\mu$m} & \colhead{60 $\mu$m} & \colhead{100 $\mu$m} \\
\colhead{ID} & \colhead{(mJy)} & \colhead{(mJy)} & \colhead{(mJy)} & \colhead{(mJy)} & \colhead{(mJy)} & \colhead{(mJy)} & \colhead{(mJy)} & \colhead{(mJy)} & \colhead{(mJy)} & \colhead{(mJy)} & \colhead{(mJy)} & \colhead{(mJy)} & \colhead{(mJy)}
}
\startdata
  S1  &  $...$  &  0.043 $\pm$ 0.003  &  0.067 $\pm$ 0.004  &  0.137 $\pm$ 0.005  &  0.467 $\pm$ 0.022  &  0.685 $\pm$ 0.025  &  $...$  &  $...$  &  $...$  &  $...$  &  $...$  &  $...$  &  $...$ \\
  S2  &  0.642 $\pm$ 0.096  &  5.134 $\pm$ 0.133  &  9.734 $\pm$ 0.224  &  13.713 $\pm$ 0.274  &  18.487 $\pm$ 0.721  &  24.303 $\pm$ 0.753  &  $...$  &  $...$  &  $...$  &  $...$  &  $...$  &  $...$  &  $...$ \\
  S3  &  0.136 $\pm$ 0.020  &  6.977 $\pm$ 0.230  &  16.607 $\pm$ 0.465  &  42.573 $\pm$ 0.936  &  83.035 $\pm$ 3.404  &  151.65 $\pm$ 4.853  &  563 $\pm$ 56  &  1287 $\pm$ 193  &  2067 $\pm$ 413  &  871 $\pm$ 130  &  2610 $\pm$ 391  &  10700 $\pm$ 1605  &  15900 $\pm$ 2385 \\
  S4  &  $...$  &  0.373 $\pm$ 0.003  &  0.737 $\pm$ 0.004  &  1.162 $\pm$ 0.006  &  1.450 $\pm$ 0.065  &  1.774 $\pm$ 0.071  &  $...$  &  $...$  &  $...$  &  $...$  &  $...$  &  $...$  &  $...$ \\
\enddata
\end{deluxetable*}
\end{longrotatetable}

\begin{deluxetable*}{lLLLL}
\tablecaption{Physical parameters of embedded sources}
\footnotesize
\label{tab:sed_par}

\tablehead{\colhead{Parameter} & \colhead{S1} & \colhead{S2} & \colhead{S3} & \colhead{S4} 
           }
\startdata
Age ($t$) [$10^{4}$~yr]  & 7.41 \pm 1.78 &  3.60 \pm 1.20 &  19.3 \pm 6.3  &   1.30 \pm 0.54  \\
Stellar mass ($M_{*}$) [$M_\sun$]  &  0.118 \pm 0.017  &  5.774 \pm 1.727  &  6.492 \pm  1.469  &  0.162 \pm 0.013  \\
Stellar radius ($R_{*}$) [$R_\sun$]  &  2.50 \pm 0.48  &  24.65 \pm 6.66  &  17.35 \pm 0.46  &  4.21 \pm 1.70  \\
Stellar temperature ($T_{*}$) [K]  &  2833 \pm 246  &  4466 \pm 875  &  6255 \pm 939  &  2931 \pm 288  \\
Envelope accretion rate ($\dot{M}_{env}$) [$10^{-5} M_\sun$~yr$^{-1}$]  &  3.002 \pm 1.219 &  10.89 \pm 2.73 & 7.026 \pm 0.292 &  0.1077 \pm 0.0369 \\
Envelope radius ($R^{max}_{env}$) [$10^3$~AU]  &  1.292 \pm 0.461  &  3.917 \pm 0.656 &  71.44 \pm 3.01  & 1.732 \pm 0.738 \\
Envelope cavity angle ($\theta_{\rm cavity}$) [deg]  &  31.19 \pm 5.52  & 20.60 \pm 1.98  &  45.19 \pm 1.07  &  20.62 \pm 3.84  \\
Disk mass ($M_{disk}$) [$M_\sun$]  &  (2.350 $\pm$ 0.335)$\times$10$^{-3}$  &  (3.055 $\pm$ 0.152) $\times$10$^{-1}$ &  (1.564 $\pm$ 0.015)$\times$10$^{-2}$  &  (2.038 $\pm$ 0.736)$\times$10$^{-4}$  \\
Disk outer radius ($R^{\rm max}_{disk}$) [AU]  &  46.96 \pm 17.67  &  113.4 \pm 27.92  &  
690 \pm  12 &  16.96 \pm 2.36  \\
Disk inner radius ($R^{\rm min}_{disk}$) [AU]  &  0.039 \pm 0.006  &  1.033 \pm 0.153  &  24.5 \pm 0.150  &  3.21 \pm 0.425 \\
Disk accretion rate ($\dot{M}_{disk}$) [$M_\sun$~yr$^{-1}$]  &  (2.429 $\pm$ 0.568)$\times$10$^{-8}$  &  (3.593 $\pm$ 0.185)$\times$10$^{-6}$  &  (1.426 $\pm$ 0.311)$\times$10$^{-7}$  &  (2.427 $\pm$ 0.369)$\times$10$^{-9}$  \\
Total luminosity ($L_{tot}$) [$L_\odot$]  &  0.39 \pm 0.13  &  240 \pm 69  &  415 \pm 7  &  1.18 \pm 0.43 \\
\hline
\enddata
\end{deluxetable*}

The SEDs of the YSOs suggest stellar infancy, with ages less than 0.2~Myr. The stellar masses range from $\sim 0.1 M_\sun$ to $6.5 M_\sun$, radii from $\sim 2.5 R_\sun$ to $24.6 R_\sun$, and temperatures from $\sim 2830$~K to 6260 K.  
The envelope and disk parameters serve to probe the evolutionary phase. Class~I objects typify earlier stages of star formation compared to Class II objects. At the initial phases, the envelope and disk accretion rates are usually high, and then decrease with time. It is expected that envelopes with accretion rates $\lesssim 10^{-6} M_\sun$~yr$^{-1}$ do not contribute significantly to the SED \citep{rob07}.
Among the four objects, S4 has the lowest envelope accretion rate 
($\sim 10^{-6} M_\sun$~yr$^{-1}$), the lowest disk mass ($\sim 10^{-4} M_\sun$), and the lowest disk accretion rate ($\sim 10^{-9} M_\sun$~yr$^{-1}$), all consistent with our earlier results of S4 being a Class~II object. The contribution to the SED from the disk related to that from the envelope is difficult to disentangle, because the envelope may not be largely dispersed yet.  The YSO S3 is associated with IRAS\,05327$+$3404 (nickname ``Holoea"\footnote{In Hawaiian for ``flowing gas"}), first detected by \citet{mag96} and  classified as being transitional between Class I and Class II, and in the process of  becoming optically exposed \citep{mag99a,mag99b}. This object, with strong far-infrared fluxes, has a very prominent circumstellar disk, extending from an inner radius of $\sim 25$~AU to an outer radius of $\sim 690$~AU.

\section{Discussion}
 \label{sec:dis}

\subsection{Molecular Cloud Morphology and Physical Parameters}
 \label{ssec:mor_par}

Despite being the most abundant molecular species in cold interstellar media, H$_{2}$ molecules are difficult to detect due to a lack of a permanent electric dipole moment. CO lines are therefore often used to trace molecular clouds.  Using $J$=1--0 $^{12}$CO and $^{13}$CO data, we map the molecular cloud structures and estimate their kinematic and physical properties. A compact and massive core is detected around the position $\alpha_{2000}$ = $84\fdg0308$, $\delta_{2000}$ = $+34\fdg1094$ for which we investigated the intensity, excitation temperature, H$_2$ column density, and velocity distribution.  The complete procedures and expressions used to derive these parameters are detailed in \citet{sun20}.

\subsubsection{Intensity and Excitation Temperature}
 \label{sssec:int_exc}

\begin{figure*}
  \centering
  \includegraphics[width=\textwidth,angle=0]{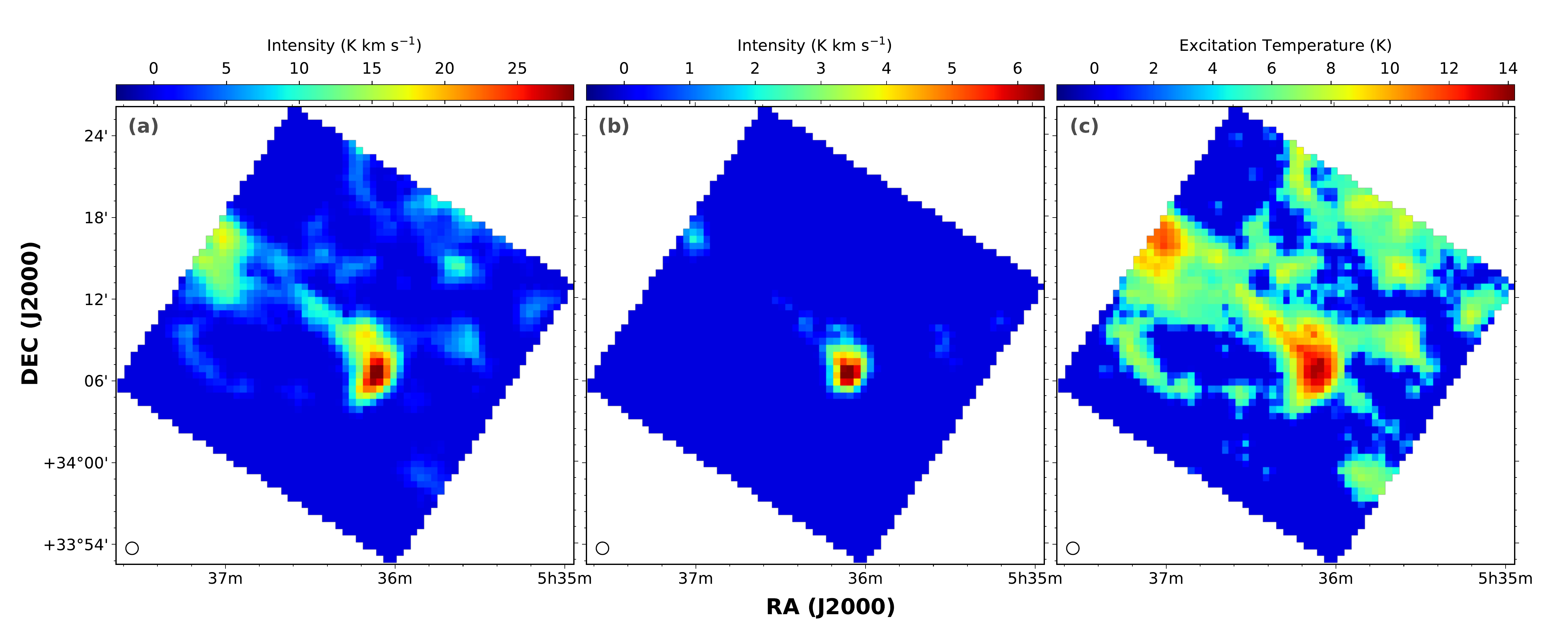}
  \caption{(a)-(b) Intensity maps of the molecular gas traced by $^{12}$CO and $^{13}$CO emission. (c) Excitation temperature map derived from $^{12}$CO by assuming optically thick. A small circle at the bottom left represents the beam size.}
  \label{fig:int_exc}
\end{figure*}

The total integrated intensity ($I_{\mathrm{CO}}$) maps for $^{12}$CO ($J$=1--0) and $^{13}$CO ($J$=1--0), along with the excitation temperature are exhibited in Figures~\ref{fig:int_exc}. The mean intensity of the cloud is calculated to be 8.8~K~km~s$^{-1}$ using $^{12}$CO and 2.6~K~km~s$^{-1}$ using $^{13}$CO. A peak in the intensity is observed at ($\alpha_{2000}$ = $84\fdg0308$, $\delta_{2000}$ = $+ 34\fdg1094$) for both isotopologues. 

Assuming $^{12}$CO is optically thick, we determined the excitation temperature ($T_{\mathrm{ex}}$) for each pixel from the peak intensity of $^{12}$CO \citep{sun20}. The excitation temperature within the cloud core ranges from 4~K to 14~K with a mean of  $7 \pm 0.6$~K.

\subsubsection{Radial Velocity}
 \label{sssec:rad_vel}

\begin{figure*}
\centering
  \includegraphics[width=0.9\textwidth]{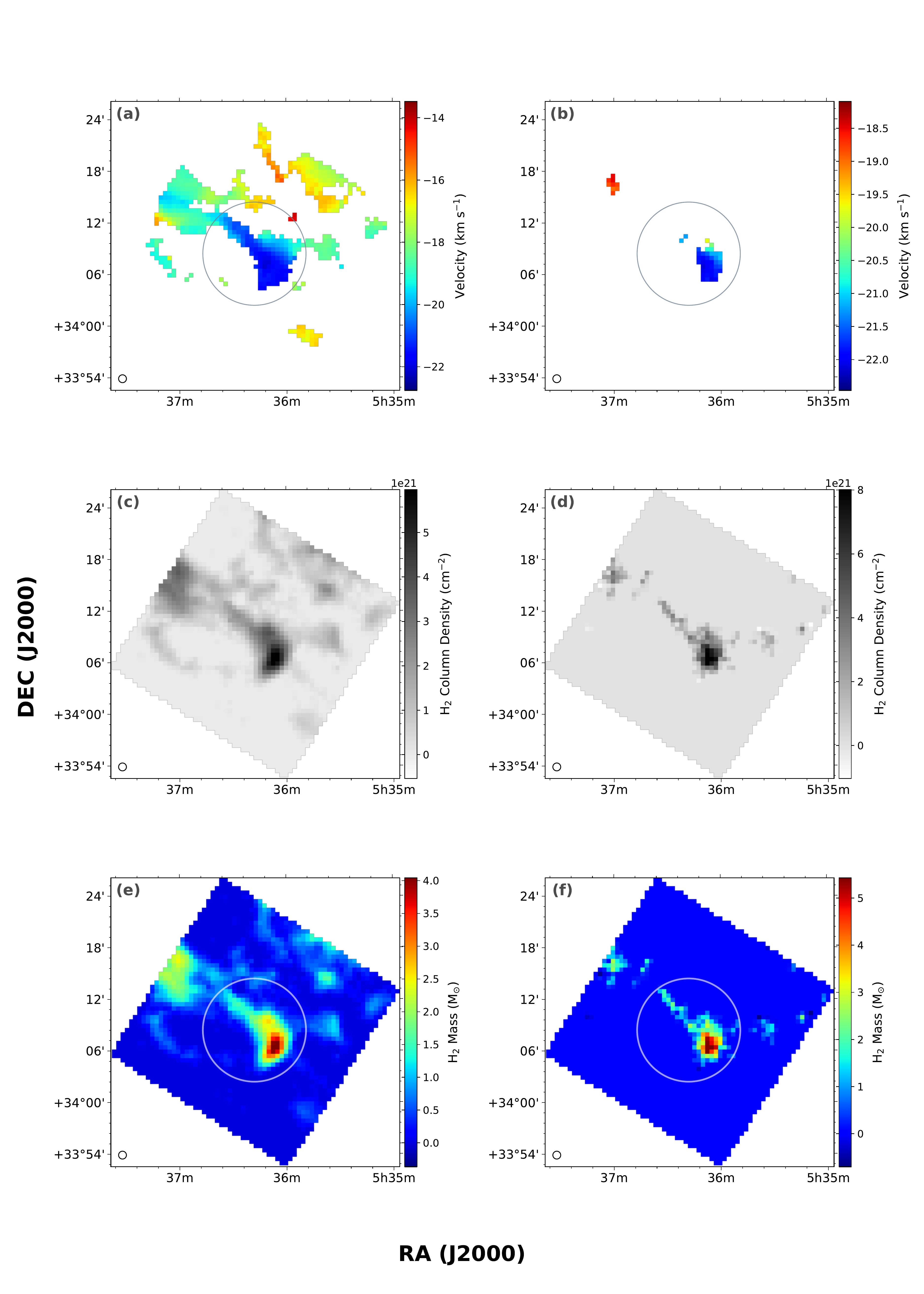}
  \caption{The velocity distributions of the emission peaks of (a)~$^{12}$CO and (b)~$^{13}$CO.   The H$_{2}$ column density distributions derived (c)~from $^{12}$CO by adopting an X-factor of $2\times10^{20}$, and (d)~from $^{13}$CO by assuming LTE. The H$_{2}$ mass distributions derived from (e)~$^{12}$CO by the X-factor method, and (f)~from $^{13}$CO by assuming LTE, by integrating the column density over the velocity channels of the clouds. 
  The circle has a radius of $6\arcmin$ from the cluster center ($\alpha_{2000}$ = $84\fdg0750$, $\delta_{2000}$ = $+ 34\fdg1400$). A small circle at the bottom left in each panel represents the beam size.}
  \label{fig:par_mul}
\end{figure*}

\begin{figure}
  \centering
  \includegraphics[width=\columnwidth,angle=0]{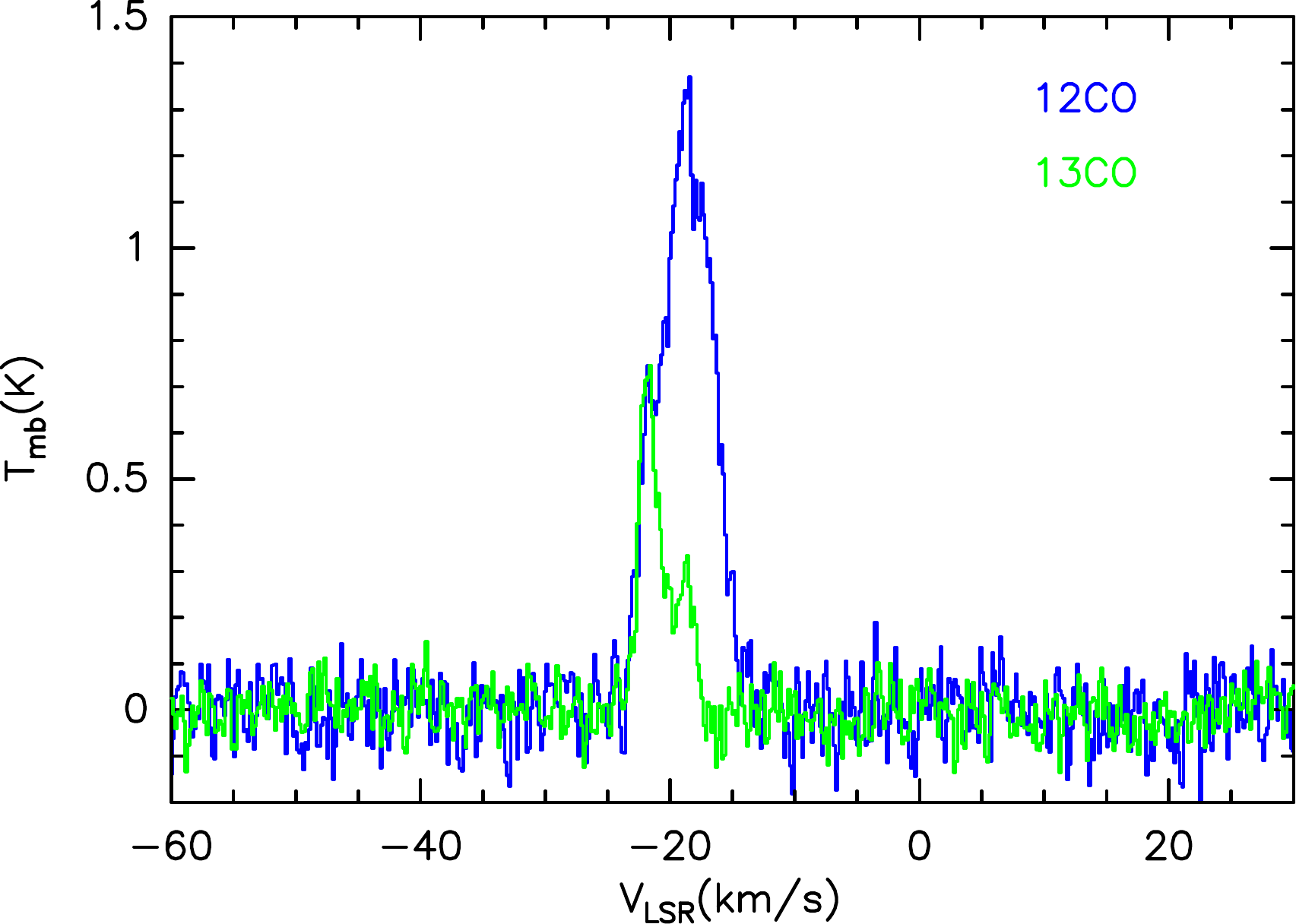}
  \caption{Averaged spectra integrated over the whole region.  Only pixels with at least three contiguous channels above $3\sigma$ are averaged.}
  \label{fig:co_spe}
\end{figure}

The velocity distribution of the cloud in $^{12}$CO and $^{13}$CO emission is shown in Figures~\ref{fig:par_mul}. Around the cloud core, the velocity structure shows a uniform distribution and varies between $-20$ and $-22$~km~s$^{-1}$. The cloud is found to be confined within a radius of $6\arcmin$ from the cluster center $\alpha_{2000}$ = $84\fdg0750$, $\delta_{2000}$ = $+ 34\fdg1400$ ($\ell = 174\fdg5345$; $b = 1\fdg0721$).  
Figure~\ref{fig:co_spe} presents the CO spectra integrated over the whole region, indicating an average velocity of $-20.2 \pm 1.7$~km~s$^{-1}$ for $^{12}$CO and $-21.5 \pm 0.7$~km~s$^{-1}$ from $^{13}$CO.

\subsubsection{Molecular Column Density}
 \label{sssec:col_den}

The H$_2$ column density ($N_{\mathrm{H_2}}$) is estimated by two methods. For $^{12}$CO, we adopted a CO-to-H$_{2}$ conversion factor, $X_{\mathrm{CO}} \equiv N_{\mathrm{H_2}}/I_{\mathrm{CO}} \equiv 2~\times~10^{20}~{\rm cm}^{-2}~({\rm K~km~s}^{-1})^{-1}$ \citep{bol13}, i.e., with the X-factor method; for the area traced by $^{13}$CO, the local thermodynamic equilibrium (LTE) method is applied \citep{sun20}. 
The H$_2$ column density in the cloud core varies, for $^{12}$CO from $2 \times 10^{20}$ to $7 \times~10^{21}$~cm$^{-2}$, with an average value of $1.5 \times 10^{21}$~cm$^{-2}$, and for $^{13}$CO from $3 \times 10^{20}$ to $9 \times 10^{21}$~cm$^{-2}$, with an average value of $3.7 \times 10^{21}$~cm$^{-2}$. The column density maps for $^{12}$CO and $^{13}$CO isotopologues are displayed in Figures~\ref{fig:par_mul}(c) for $^{12}$CO, and \ref{fig:par_mul}(d) for $^{13}$CO.

\subsubsection{H$_{2}$ Mass}
 \label{sssec:h2_mass}

The molecular mass is estimated by integrating the column density over the area of each region. To calculate the H$_{2}$ mass, again we used the X-factor method for $^{12}$CO, and the LTE method for $^{13}$CO, and the mass distributions are presented in Figure~\ref{fig:par_mul}(e) and 
Figure~\ref{fig:par_mul}(f). A compact cloud structure is detected within a radius of 6$\arcmin$ from the cluster center traced by both $^{12}$CO and $^{13}$CO emission. The total mass of this cloud complex is estimated to be $313 \pm 0.9$~M$_\sun$ for $^{12}$CO and $210 \pm 1.2$~M$_\sun$ for $^{13}$CO. 
Generally $^{12}$CO emission traces the total gas content distributed in the molecular cloud, including lower density ($\sim 10^2$~cm$^{-3}$) diffuse gas, while the optically thinner $^{13}$CO emission traces typically denser ($\sim 10^{3}$--$10^{4}$~cm$^{-3}$) components, a reason of higher H$_{2}$ mass derived from $^{12}$CO compared with $^{13}$CO.

\subsection{Physical Association Between the YSOs and Molecular Gas}
 \label{ssec:mol_yso}

\begin{figure*}
\centering
  \includegraphics[width=\textwidth]{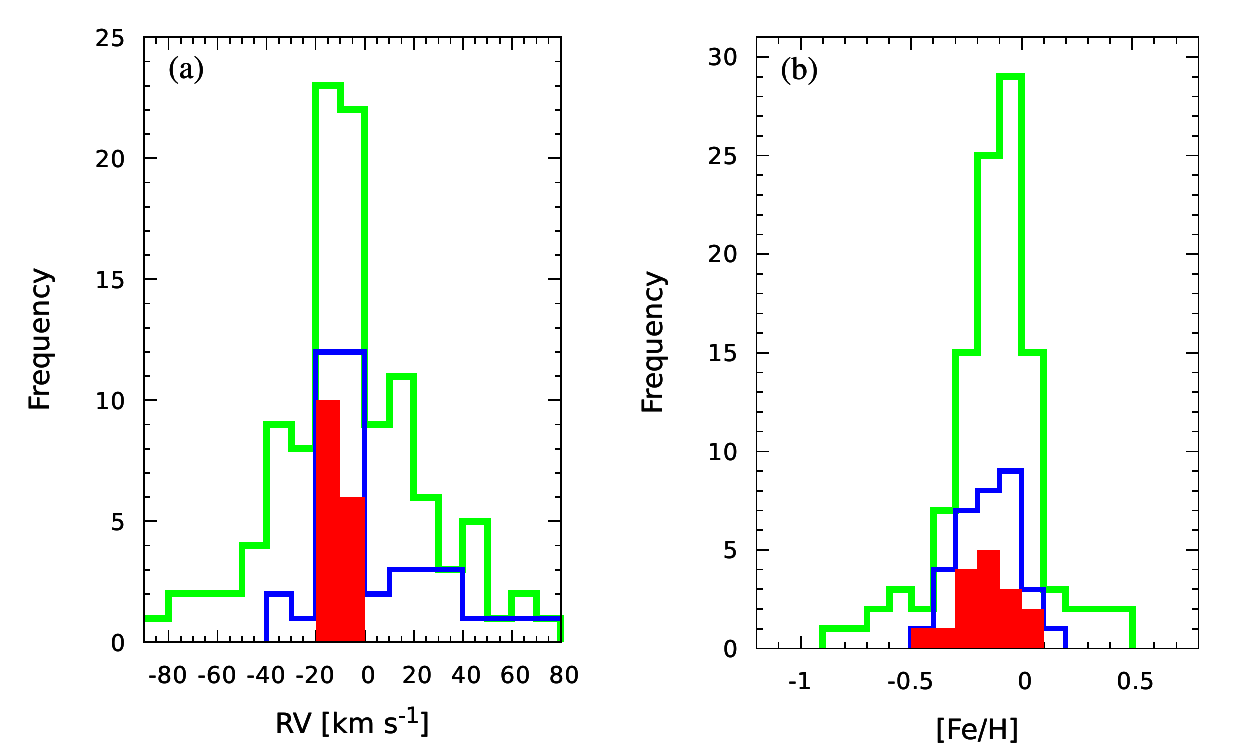}
  \caption{(a)~Radial velocity (RV) and (b)~metallicity ([Fe/H]) distributions of the sources from the LAMOST DR5 data.  The color scheme for the histograms is the same as in Figure~\ref{fig:cmd_mul}, i.e., sources satisfying the parallax criteria only are marked in green, those satisfying the proper motion criteria are in blue, and those satisfying both 
  (member candidates) are in red.}
  \label{fig:lam_mul}
\end{figure*}

Of the 200 member candidates of M\,36, 16 (8\%) were observed as LAMOST~DR5 targets.  RV measurements were not included in our membership selection, because of the limited amounts of data available, and because of possible variations due to binary orbital motion. The RV and metallicity measurements are presented in Figure~\ref{fig:lam_mul}.  The RVs of the majority of members range between $-5$~km~s$^{-1}$ and $-20$~km~s$^{-1}$.  This is consistent with that reported by \citet[][RV$=-17.83 \pm 0.99$~km~s$^{-1}$]{fri08}, and matches approximately 
with the RV ($\approx -21$~km~s$^{-1}$) of the molecular cloud.  

The distance to the cloud is uncertain.  Adopting RV$=-21$~km~s$^{-1}$, the nominal Milky Way rotation \citep{rei19} suggests a kinematic distance of $1.68 \pm 0.05$~kpc, placing the cloud somewhat in the background but physical association of the cloud with the cluster ($\sim1.2$~kpc) cannot be ruled out, in evidence notably of the {\it Gaia} distance of $1.395_{-0.26}^{+0.40}$~kpc (Table~\ref{tab:ysos_phot}) for the young object S2 in the cloud.  While not unambiguously established at this time, the proximate angular separation, radial velocity, and distance provides tantalizing evidence of a physical connection between the cloud and the cluster.

The metallicity distribution of member candidates peaks around $-0.1$ to $-0.2$, with an average of $-0.15 \pm 0.10$.  Though the dispersion is relatively large, our result, based on a relatively reliable list of members, suggests a subsolar metallicity for M\,36.

Star formation activity at the earliest stages (1--2 Myr) is difficult to trace due to heavy dust obscuration. Star formation is believed to be regulated by dense gas in molecular clouds \citep{gao04,wu05, dut18}, and the star formation rate is known to be correlated with the molecular mass \citep{won02, lad10, lad12}. 
Surveys of star-forming regions have demonstrated that approximately 75\% of the stars are formed in groups or clusters \citep{car00, all07, gut09}, whereas about 80\% of all young stars are located in embedded clusters \citep{lad03, por03}. Several factors are involved in the formation and early evolution of a cluster, such as the structure of the parental molecular cloud \citep{sam15}, fragmentation in the parental cloud due to turbulent motions and/or gravity, dynamical motions of the young stars, and the feedback from young stars \citep{gut09}.
While infrared surveys are an eminent tool to parameterize young embedded clusters, data at much longer millimeter wavelengths are also powerful to trace the gaseous contents and the distribution of dust throughout a molecular cloud. 

\begin{figure*}
\centering
  \includegraphics[width=\textwidth]{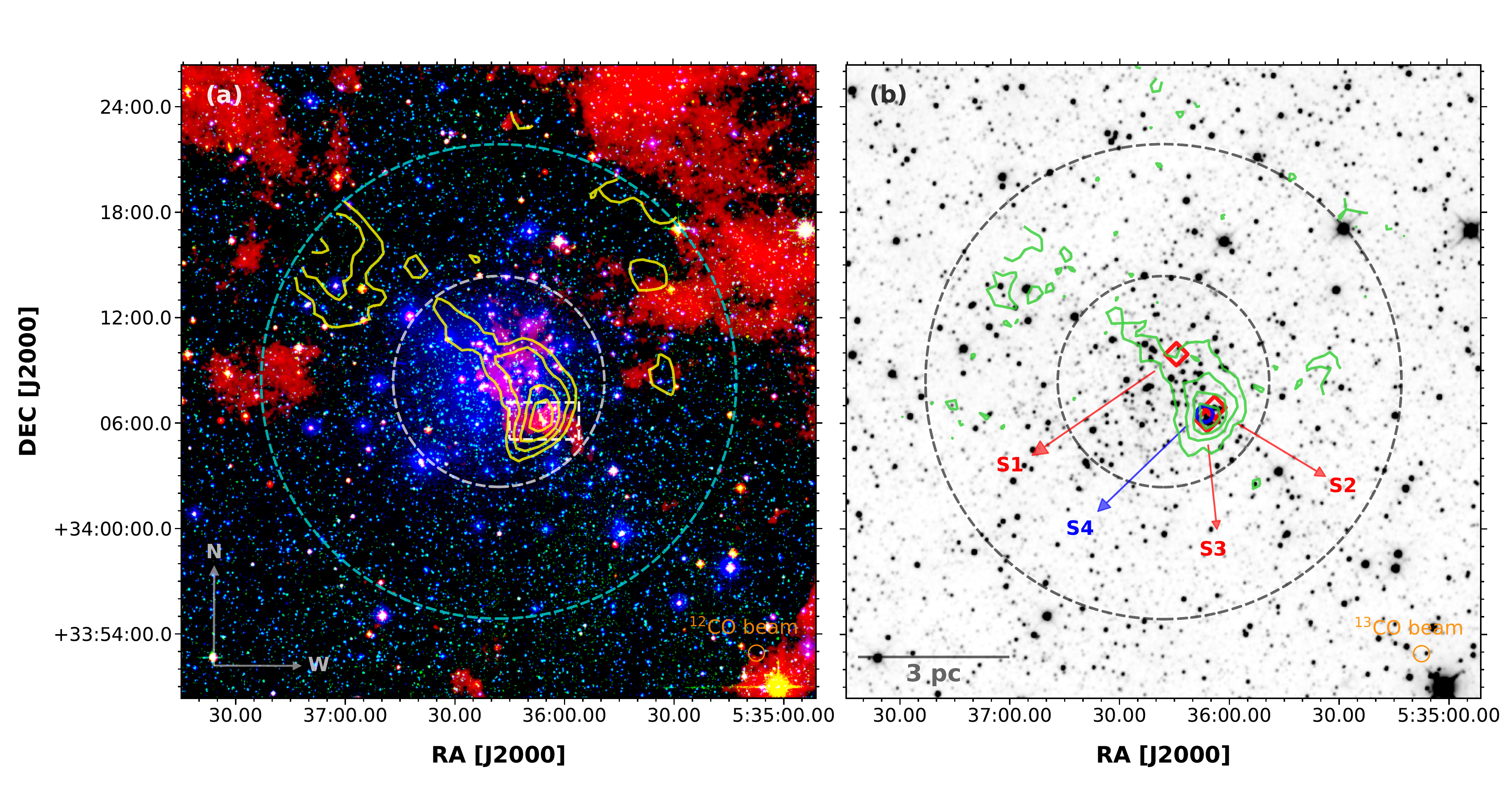}
  \caption{(a)~Color composite image of M\,36 using optical and infrared data, taken from DSS2~$B$~0.44~$\mu$m (blue), 2MASS~$K$~2.2~$\mu$m (green), and {\it WISE}~$W3$~12~$\mu$m (red) for a region of $30 \arcmin\times30\arcmin$. The $^{12}$CO ($J$=1--0) distribution is traced by the yellow contours, at contour levels of (6.27, 12.55, 18.82, 25.09, and 31.37)~K~km~s$^{-1}$. The high extinction complex (Section~\ref{ssec:ext_map}) is depicted by the white rectangle. 
  (b)~The corresponding {\it WISE}~$W2$~4.6~$\mu$m image, with the green contours representing the $^{13}$CO ($J$=1--0) integrated intensity, at the contour levels at (1.62, 3.24, 4.86, 6.48, and 8.10)~K~km~s$^{-1}$. The four YSOs (Section~\ref{ssec:yso_iden}) are indicated. The circles of radii of $6\arcmin$ and $13\farcm5$ mark the possible boundaries of the molecular cloud and of the cluster extension, respectively.}
  \label{fig:co_mol}
\end{figure*}

A color composite image of the M\,36 cluster, made from optical ($B$ band), near-infrared ($K$ band), and far-infrared ({\it WISE/W3} 12~\micron) data, is presented as Figure~\ref{fig:co_mol}(a). The {\it WISE} 12~$\mu$m emission, sensitive to warm dust \citep{dun08}, coincides well with the extinction complex and CO gas.  The $^{12}$CO and $^{13}$CO distributions reveal a compact cloud with a linear extent of $\sim 10\farcm4$ (3.6~pc), extending from north-east to south-west plus a core located at the south-west corner where the three YSOs discussed earlier reside near the densest part, within $\sim 1\arcmin$, of the cloud core.  This is where active star formation is taking place.

\subsection{Sustaining Star Formation}
 \label{ssec:sus_sta}

The timescale of star formation plays a significant role in the evolution of a star cluster. It depends not only on the initial cloud conditions, such as the density and temperature profiles, but also on how the new-born stars interplay with the remnant clouds, so as to induce the birth of next-generation stars, e.g., by the dynamical compression of an \ion{H}{2} front, and change in self-gravity \citep{fuk00}.   Alternative to clustering, star formation may proceed in an isolated or distributed mode \citep{koe08, eva09}, leading to multiple stellar populations within a cloud \citep{gra12, bou15, bek17}.

The four YSOs identified by this work are in the proximity of the cloud core, with angular separations of $3\farcm69$ (for S1), $0\farcm79$ (for S2), $0\farcm46$ (for S3), and $0\farcm15$ (for S4).  The source S2 has a measured {\it Gaia}~DR2 distance of $1.395_{-0.26}^{+0.40}$~kpc (Table~\ref{tab:ysos_phot}), matching well with the cluster distance.  In particular, S3 coincides spatially with an IRAS source (IRAS~05327$+$3404), which drives a powerful ionizing outflow with a velocity of $\sim 650$~km~s$^{-1}$, that aligns with a CO outflow \citep{mag96}.  Its SED indicates stellar infancy with ample circumstellar material.  S3 is clearly a part of the continuing stellar formation episode in the dense core, and its outflows may have a profound effect on the surrounding environments to prompt future formation of stars.  

Star formation processes in an open cluster may span about 5~Myr to 20~Myr \citep{lim16}, with the low-mass members spending up to some $\sim10^8$~years in the pre-main sequence phase \citep{her62}.  
Star clusters older than $\sim 5$~Myr are typified with little parental molecular gas \citep{lei89}.  
M\,36, with an age of $\sim15$~Myr (Section~\ref{ssec:cmd_ana}), is therefore unusual in connection with a nearby voluminous molecular cloud harboring a protostellar population with ages $< 0.2$~Myr.  
More evidence is clearly needed to confirm or to discredit the physical connection between the star-forming cloud and the cluster.  If the cloud and the cluster share the same formation scenario, this presents a rare case of sustaining star formation in a cloud complex.  Even if the cloud happens to be in the intermediate background of the cluster, our discovery of a cloud active in star formation offers added information of the star formation history in this part of the Galactic disk.

\section{Summary and Conclusions}
 \label{sec:con}

We report the stellar contents of the nearby young open cluster M\,36 using multiwavelength data sets.  Cluster membership is diagnosed using five dimensional astrometric parameters. The nature of the young objects and associated molecular cloud is analysed using molecular emission ($^{12}$CO and $^{13}$CO lines), supplemented with infrared photometry. The key outcomes of this work are summarized as follows:

\begin{enumerate}

 \item A list of 200 member candidates are presented based on proper motions and parallax measurements from the {\it Gaia}~DR2 catalog.  Applying the same set of selection criteria on a nearby control field, a false positive rate of 8\% is expected.

 \item The cluster exhibits a distinct proper motion from the field, with member candidates concentrated within 0.5~mas~yr$^{-1}$ around the peak of $\mu_{\alpha}$ $cos\delta = -0.15 \pm  0.01$~mas~yr$^{-1}$, and $\mu_{\delta} = -3.35 \pm 0.02$~mas~yr$^{-1}$.  The parallax measurements of the member candidates suggest a parallax range of 0.7--0.9~mas with a median of $0.82 \pm 0.07$~mas (distance $\sim1.20 \pm 0.13$~kpc).  The cluster has an angular diameter of $27 \arcmin \pm 0\farcm4$, equivalent to a linear extent of $9.42 \pm 0.14$~pc.  The member candidates have an age of $\sim15$~Myr, consistent with the literature values. 

 \item The $K$-band extinction map leads to identification of a high extinction ($A_{V} \sim 23$~mag), small complex ($\sim 1\farcm9 \times 1\farcm2$).  This high-extinction region coincides with a molecular cloud core, both in $^{12}$CO and $^{13}$CO.  The cloud shows a uniform velocity ($-20$~to~$-22$~km~s$^{-1}$) structure with a total mass of (2--3)$\times10^{2}$~M$_{\sun}$. In addition to spatial association, the cloud's radial velocity agrees with that of the cluster members, indicative of physical association.  Four protostars with ages $< 0.2$~Myr are found, with three of them being located in close proximity to the high extinction complex. 
 
 \item The discovery of a molecular cloud harboring stars in their infancy provides new clues on the star formation history in the vicinity of M\,36.  If the cloud is physically associated with the cluster, this presents a tantalizing case of sustaining starbirth in a cloud complex, with a cluster formed $\sim15$~Myr ago, to a dense molecular cloud core currently active in star formation.

\end{enumerate}

\section*{Acknowledgements}

AP and SD acknowledge the support by S. N. Bose National Centre for Basic Sciences, funded by the Department of Science and Technology, India, and by  the Taiwan Experience Education Program (TEEP) during their visit to National Central University.  
YG's research is supported by the National Key Basic Research and Development Program of China (2017YFA0402700), the National Natural Science Foundation of China (U1731237, 12033004), and Chinese Academy of Sciences Key Research Program of Frontier Sciences (QYZDJSSW-SLH008). 
This work has made use of data from the European Space Agency (ESA) mission {\it Gaia} (\url{https://www.cosmos.esa.int/gaia}), processed by the {\it Gaia} Data Processing and Analysis Consortium (DPAC, \url{https://www.cosmos.esa.int/web/gaia/dpac/consortium}). 
This study used the data from UKIDSS, which is a next generation near-infrared sky survey using the WFCAM on the United Kingdom Infrared Telescope on Mauna Kea in Hawaii. This work made use of data products from the 2MASS, which is a joint project of the University of Massachusetts and the IPAC/Caltech. 
This work also used observations made with the {\it Spitzer Space Telescope}, operated by the Jet Propulsion Laboratory, Caltech under a contract with NASA. This work used the data products from LAMOST, operated and managed by the National Astronomical Observatories, CAS.
The authors are thankful to the anonymous referee for providing constructive suggestions that significantly improved the scientific content of the paper. 

\software{GILDAS \citep{pet05}}

\end{document}